# Genetic recombination is targeted towards gene promoter regions in dogs

Authors: Adam Auton (1), Ying Rui Li (2), Jeffrey Kidd (3), Kyle Oliveira (4), Julie Nadel (1), J. Kim Holloway (4), Jessica J. Hayward (4), Paula E. Cohen (4), John M. Greally (1), Jun Wang (2, *), Carlos D. Bustamante (5, *), Adam R. Boyko (4, *)

1. Department of Genetics, Albert Einstein College of Medicine, 1301 Morris Park Ave, Bronx, NY 10461, USA
2. BGI-Shenzhen, Shenzhen 518083, China
3. Department of Human Genetics, University of Michigan, Ann Arbor, Michigan, USA
4. Department of Biomedical Sciences, Cornell University College of Veterinary Medicine, Ithaca, NY 14853, USA
5. Department of Genetics, Stanford University, Stanford, CA 94305, USA

Corresponding Authors:

wangj@genomics.org.cn (J. W.)
cdbustam@stanford.edu (C. D. B)
boyko@cornell.edu (A. R. B.)

## Abstract

The identification of the H3K4 trimethylase, PRDM9, as the gene responsible for recombination hotspot localization has provided considerable insight into the mechanisms by which recombination is initiated in mammals. However, uniquely amongst mammals, canids appear to lack a functional version of *PRDM9* and may therefore provide a model for understanding recombination that occurs in the absence of *PRDM9*, and thus how *PRDM9* functions to shape the recombination landscape. We have constructed a fine-scale genetic map from patterns of linkage disequilibrium assessed using high-throughput sequence data from 51 free-ranging dogs, *Canis lupus familiaris*. While broad-scale properties of recombination appear similar to other mammalian species, our fine-scale estimates indicate that canine highly elevated recombination rates are observed in the vicinity of CpG rich regions including gene promoter regions, but show little association with H3K4 trimethylation marks identified in spermatocytes. By comparison to genomic data from the Andean fox, *Lycalopex culpaeus*, we show that biased gene conversion is a plausible mechanism by which the high CpG content of the dog genome could have occurred.

## Author Summary

Recombination in mammalian genomes tends to occur within highly localized regions known as recombination hotspots. These hotspots appear to be a ubiquitous feature of mammalian genomes, but tend to not be shared between closely related species despite high levels of DNA sequence similarity. This disparity has been largely explained by the discovery of *PRDM9* as the gene responsible for localizing recombination hotspots via recognition and binding to specific DNA motifs. Variation within *PRDM9* can lead to changes to the recognized motif, and hence changes to the location of recombination hotspots thought the genome. Multiple studies have shown that *PRDM9* is under strong selective pressure, apparently leading to a rapid turnover of hotspot locations between species. However, uniquely amongst mammals, *PRDM9* appears to be dysfunctional in dogs and other canids. In this paper, we investigate the how the loss of *PRDM9* has effected the fine-scale recombination landscape in dogs and contrast this with patterns seen in other species.

## Introduction

Until recently the mechanisms controlling the localization of recombination hotspots in mammalian genomes were largely unknown. However, recent research has revealed that zinc-finger protein, PRDM9, binds to specific DNA motifs in the early stages of recombination initiation in order to direct such events [1-3]. PRDM9 trimethylates lysine 4 of histone H3 (H3K4me3), an epigenetic modification specifically enriched around recombination initiation sites [4-6]. The importance of PRDM9 for the localization of recombination events has been demonstrated in both humans [7,8] and mice [2,6], with recent results in mice suggesting that PRDM9 determines the location of virtually all recombination hotspots in these organisms [6].

Variation in the zinc-finger encoding domain of *PRDM9* in humans can alter the DNA motif to which the protein binds, and in turn alter the activity of recombination hotspots [7-9]. High levels of variation in *PRDM9* across species [2,10,11] may explain why humans and chimpanzees do not share recombination hotspots despite very high levels of overall sequence identity [10]. Indeed, *PRDM9* shows clear evidence of rapid evolution across the metazoan taxa covering an era of roughly 700 million years [11]. Despite successful formation of double-strand breaks (DSBs) [6], the initiating event in meiotic recombination, knockout of *Prdm9* results in infertility in both male and female mice with arrest of spermatogenesis and oogenesis at pachynema, impairment of DSB repair, chromosome asynapsis, and disrupted sex-body formation in males [12]. Intriguingly, *Prdm9* has been shown to be involved with hybrid sterility, potentially implicating it in the process of speciation [12].

Given the wide-ranging importance of *PRDM9*, it was therefore surprising to note that dogs (*Canis familiaris*) and other canids are the only known mammals to carry functionally inert versions of *PRDM9* with multiple disruptive mutations [11,13]. This implies that either the function of *PRDM9* is carried out by another gene, or that dogs have been able to avoid the loss of fertility associated with loss of *PRDM9* while also ensuring that recombination continues to occur.

In order to gain insight into recombination across the canine genome, we have constructed a genetic map using patterns of linkage disequilibrium (LD) estimated from next-generation sequencing of 51 dogs. Methods for estimating recombination rates from polymorphism data have been validated at both broad and fine scales [14-16], and have previously been used to obtain relatively broad-scale recombination rate estimates in dogs via the use of SNP microarray data [13]. A potential concern when using such methods in dogs is that breed-formation bottlenecks can lead to considerable levels of inbreeding. For this reason, we have utilized genetic polymorphism data collected primarily from free-ranging dogs from geographically diverse regions (Supplementary Table 1) and largely lacking a history of excessive selective breeding following the original domestication event [17]. These non-breed dogs, which we term 'village dogs', show dramatically reduced levels of homozygosity, and a faster decay of LD when compared to inbred breed dogs (Supplementary Figure 1).

## Results

The dogs were sequenced to 8-12X coverage with 101bp paired end reads (Supplementary Table 2), allowing identification of 13.6 million autosomal variants, and 366,000 variants on chromosome X. Based on comparisons to Illumina CanineHD microarray [18] SNP data, we estimate that we have >98% power to detect variants with a minor allele frequency of 5%, and a genotype accuracy of 99.1%. As estimation of genetic maps can be moderately sensitive to false-positive variant calls [10], we performed extensive variant filtering to identify a subset of high-quality variants (see Methods). Our filtered set consisted of 3.5 million autosomal SNPs, and 198,000 SNPs on the X chromosome, which we used to construct the genetic map.

To validate our recombination rate estimates, we compared our estimates to broad-scale experimental estimates obtained from pedigree studies [19]. There is strong agreement between our map and the linkage map at the broad scale (Pearson r=0.87 at 5Mb; Figure 1A, Supplementary Figure 2, and Supplementary Figure 3). Consistent with observations in other species [20], recombination rates tend to be highest in telomeric regions, and lowest near the centromere (Figure 1B). The correlation between chromosome physical length and total map length is similar for our map and the linkage map (Pearson r = 0.88 and r = 0.83 respectively; Supplementary Figure 4). We conclude that our recombination rate estimates obtained from patterns of LD in population sequencing data are capable of accurately recapitulating the canine genetic map.

Our estimates suggest a more uniform distribution of recombination in the dog genome than has been seen in human (Supplementary Figure 5A), as has been reported previously [13]. However, simulations indicate that the estimate of this distribution from patterns of linkage disequilibrium is sensitive to the effective population size, with larger effective population sizes leading to higher estimates of the background recombination rate (Supplementary Figure 5B). As such, given that the effective population size estimated in dogs is larger than human, we cannot conclude the recombination is more uniformly distributed in dog.

Previous LD-based estimates of recombination in dog by Axelsson *et al*. were obtained using a microarray with 170k markers [13], with estimates averaged over multiple breeds. At the broad scale, there is good agreement between the two studies, which correlate more strongly with each other than the pedigree-based map (Supplementary Figure 6). However, the increased marker density of our study allows investigation of the fine-scale recombination landscape in dogs. Despite the apparent loss of PRDM9 in dogs, we detect 7,677 hotspot-like peaks in the recombination rate throughout the canine genome, with a median width of 4.3kb (Figure 1C).

We compared these hotspots to those identified by Axelsson *et al*. [13] and found we could confirm the presence of 1,090 out of the 4,074 hotspots identified by that study (27%). The overlap between the two sets is strongly significant ($p<0.0001$, assessed by 10,000 randomizations of the hotspot locations), suggesting both datasets are picking up on real signal. However, the relatively low concordance suggests either low power or an elevated false positive rate in one or both studies. To investigate further, we have plotted the average recombination rate as measured by the Axelsson study around hotspots identified by our study (Supplementary Figure 7A). We see that recombination rate estimates tend to be higher for hotspots identified by both studies. However, hotspots identified by our study alone still show a peak in recombination in the Axelsson study, suggesting agreement between the two studies even when there was in sufficient power to call a hotspot in the Axelsson study. Conversely, the average recombination rate as measured by our study around hotspots identified by the Axelsson study (Supplementary Figure 7B) shows only very weak elevation of recombination around hotspots not identified by our study.

### High recombination in CpG-rich regions

In humans and mice, specific DNA motifs have been implicated as the binding sites for *PRDM9* [1,4,21]. In order to investigate if DNA motifs could be identified within canine recombination hotspots, we selected 6,228 hotspot regions with no missing sequence data. For each hotspot we identified a region on the same chromosome showing no evidence for local recombination rate elevation ('coldspots'), and with GC content within 0.5% of that of the hotspot, and CpG content within 0.1%. If more than one such region could be found, we selected the one that matched the hotspot most closely in terms of SNP density. In this way, we were able to identify 4,759 hotspots with matched coldspots.

Using the sequences of the matched hotspots and coldspots, we performed a search of motifs showing enrichment in hotspot sequences. Our results indicate an extremely strong association with CpG-rich motifs (Table 1), with the most significant motif being the 7-mer CGCCGCG (p = 1.1e-21, Fisher's Exact Test after Bonferroni correction), which is found in 21.3% of hotspots but only 13.2% of coldspots, a relative enrichment of 61%. These highly CpG-rich motifs retain significantly high levels of enrichment in hotspots having masked either repeat or non-repeat DNA sequence.

Both GC and CpG content show a strong association with canine recombination at fine scales (Supplementary Figure 8A and B). However, CpG content shows a stronger correlation with recombination rate than GC content over multiple scales (r=0.37 vs r=0.25 respectively at 1Mb; Supplementary Figure 8C and D). If both measures are included as predictors in a multiple regression model, CpG content has a positive association, whereas GC content is negative (Supplementary Table 3).

The influence of GC and CpG content can also be seen when considering the average recombination rate around DNA repeats. The most recombinogenic repeats are low-complexity with high levels of GC and CpG content (Figure 2A). In contrast, the majority of LINE and SINE elements exhibit recombination rates close to the genome average, with a few such as Looper and L1_Canid2 showing weak suppression of recombination.

The association between recombination and CpG-dense regions is suggestive of an association with gene promoter regions. Indeed, we observe highly elevated rates of recombination around transcription start sites (TSS; Figure 2B and Supplementary Figure 9), dwarfing the elevation that has been observed around TSS in humans and chimps [10,22]. Of the 7,677 called hotspots, 29% overlap with a TSS, and 50% are within 14.7kb. Only a small fraction (14%) of hotspots appear to be over 100kb from a TSS. However, the elevation in recombination rate around TSS appears to be associated with CpG islands serving as promoter regions rather than the TSS themselves, as the recombination peak is shifted away from the TSS for genes with the nearest CpG island at some distance from the TSS (Figure 2C), with genes without a nearby CpG island not showing large peaks in local recombination. Conversely, CpG islands containing TSS show elevated recombination rates relative to CpG islands at some distance from TSS (Supplementary Figure 10), although interestingly CpG islands > 10kb from the nearest TSS show higher rates than those near (but not containing) a TSS.

### Recombination has increased the CpG content of the dog genome

*PRDM9* is thought to have been disrupted early in canid evolution, as previous work has shown that the amino acid coding sequence contains multiple disruptive mutations across a diverse set of canid species [11,13]. We have further investigated the extent of *PRDM9* disrupting mutations within the Canidae family by sequencing within exon containing the zinc-finger domain of *PRDM9* in the *Lycalopex* and *Urocyon* genera. Within in the Andean fox, *Lycalopex culpaeus* (6 – 7.4 Mya divergence from dogs [23]), we found the same disrupting frameshift mutation as has observed in dog (Supplementary Table 4 and Supplementary Table 5), as well as an additional frameshift, and a premature stop codon. In the Island fox, *Urocyon littoralis,* (>10 Mya divergence from dogs [23]), while we do not observe the same mutations seen in dog, we do observe a distinct premature stop codon, indicating that PRDM9 has been disrupted in this species as well. As none of the identified mutations are common to all species, it would appear that the original disruptive mutation likely occurred outside of sequenced exon.

Due to the early loss of *PRDM9*, it has been suggested that fine-scale patterns of recombination may be shared across species in the canid lineage [13], in contrast to other mammalian species in which hotspots are not shared [10]. Such inferences have been based on the effect of Biased Gene Conversion (BGC), in which a recombination-associated heteroduplex in the vicinity of an existing polymorphism can produce base-pair mismatches that are preferentially repaired with C/G alleles rather than A/T alleles (Supplementary Figure 13). As such, BGC increases the probability of a C/G allele being transmitted to the next generation, and sustained BGC can ultimately alter the base composition of the genome [24].

To investigate if BGC is active around canine recombination hotspots, we consider the ratio of AT->GC polymorphisms to GC->AT with local regions of the genome (see Supplementary Material). In order to polarize polymorphisms in dog, we sequenced a female Andean Fox, which, as described above, is diverged from dogs by approximately 6 – 7.4 million years [23] and also lacks a functional version of *PRMD9*. The fox sample was sequenced to approximately 11X coverage using 100bp paired-end Illumina sequencing. In the absence of a reference genome for this species, reads were mapped to the dog genome (canFam3.0). Using this data, we polarized the ancestral and derived alleles for polymorphisms observed in dog by assuming that shared alleles represent the ancestral allele. Using this data, we were able to polarize 3.2 million polymorphisms on the dog lineage with high confidence. Likewise, we performed SNP discovery in the fox data, and were able to identify and polarize 1.2 million polymorphisms on the fox linage.

Around canine recombination hotspots, we observe a localized skew in the rate of AT to GC acquisition in the dog genome (Figure 3A), which is stronger for hotspots with higher peak recombination rates (Figure 3B; Supplementary Figure 11A). The effect remains visible even after excluding all polymorphisms within putative CpG sites (Supplementary Figure 11B). The dog genome is notable for its high density of CpG-rich regions [25,26]. Given CpG-rich regions are highly recombinogenic in the dog genome, it is plausible that BGC is the mechanism by which the dog genome has acquired its high density of CpG rich regions. Specifically, if CpG regions are promoting recombination, and are thereby acquiring increasing levels of GC content via BGC, this would in turn further increase the CpG content of the region and hence further increase recombination in a self-reinforcing process.

If dogs and foxes shared the same hotspots, a similar pattern would be expected for polymorphisms observed on the fox linage. However, we do not see evidence of a skew in fox linage polymorphisms around canine recombination hotspots (Figure 3A), which would imply that these two species do not share recombination hotspots.

### The association between H3K4me3 and canine recombination

The role of PRDM9 is to trimethylate histone H3K4, and studies in mice have shown that nearly 95% of hotspots overlap an H3K4me3 mark [4]. It is therefore interesting to ask if canine recombination maintains an association with H3K4me3, especially given the apparent elevation of recombination around gene promoter regions in dogs. We have used ChIPseq to identify regions of H3K4me3 in dog spermatocytes during the leptotene / zygotene (L/Z) and pachytene phases of prophase I of meiosis. We identified 28,349 autosomal ChIPseq peaks in L/Z and 32,830 for pachynema. Of these, 8,721 (31%) peaks were unique to L/Z and 13,613 (41%) unique to pachynema.

While recombination rates do appear highly elevated around H3K4me3 marks, the effect is almost entirely explained by the presence or absence of a putative CpG island overlapping the mark (Figure 4A and B). The pattern is very similar for both L/Z and pachytene cells, albeit with a small increase in rate for pachytene-specific marks in the absence of CpG islands (Supplementary Figure 12). Conversely, while CpG islands overlapping H3K4me3 marks are ~60% more recombinogenic than those without marks, a strong elevation in local recombination rate remains visible for islands without H3K4me3 marks. Notably, putative CpG islands without H3K4me3 marks also appear to have elevated background rates (Figure 4C), and this effect persists even after extensive thinning CpG islands to ensure lack of clustering. This could reflect background sequence context, as islands with H3K4me3 marks have lower levels of CpG content in the flanking 50kb than those without marks (2.2% vs 1.2%, $p << 1e-16$), or may be indicative of other epigenetic factors such as DNA methylation.

## Discussion

The apparent loss of *PRDM9* in canids makes dogs of particular interest for the study of meiotic recombination. Our study reveals that while broad scale patterns of recombination appear superficially typical of mammals, dogs appear to have a quite different landscape compared to other mammals so far studied at the fine scale. Of particular note is the strong association of recombination with CpG-rich features of the genome, particularly around promoter regions, which is reminiscent of the double-strand break localization around H3K4me3 marks that has been observed around promoter regions in *Prdm9*-knockout mice [6]. However, in contrast to these results, we find that the elevation in canine recombination rate around promoters appears to be primarily associated with CpG content, and shows little association with H3K4me3 marks. The association with promoter regions is also superficially reminiscent of the elevated recombination rates observed around promoter in yeast, which has been related to nucleosome spacing [27,28].

The biased conversion of A/T to G/C alleles at already CpG enriched recombination hotspots may help explain the 2-3-fold increase in putative CpG islands in the dog genome compared to human and mouse [25]. Despite the relative high density of CpG islands in the dog genome, it has also been noted that dogs have fewer promoter-associated CpG islands than humans or mice, especially near essential and highly expressed genes [25]. If the role of *PRDM9* is to deflect DSBs away from functional elements, as has been suggested in other species [6], then the preferential loss of recombinogenic CpG islands near promoters in dogs may indicate that selection is acting to deflect recombination from genic regions since the loss of *PRDM9*.

*PRDM9*-knockout mice are infertile due to the failure to properly repair DSBs [29]. How canids escaped this infertility is unknown, but it must have occurred in the common ancestor of dogs, wolves, foxes and jackals, but after the split from Panda ~49Mya [13]. The stable recombination landscape of canids resulting from the loss of *PRMD9* may contribute to the ability to successfully hybridize relatively divergent canine species (e.g. dog and jackal [30]). Our comparison of substitution patterns in dogs and Andean foxes does not support the hypothesis that the death of *PRDM9* has resulted in the evolutionary stability of recombination along the canid lineage, at least at the fine scale. It is therefore possible that while *PRDM9* is dysfunctional, an unknown ortholog of *PRDM9* could have assumed a similar role.

Nonetheless, the observed recombination landscape in dogs does appear to have some unusual features, and it is plausible that these result directly from the loss of *PRDM9*. However, it is worth noting that in addition to the loss of *PRDM9*, canines are considerably diverged from other species that have so far been studied for fine-scale recombination rate variation, and it is plausible that the observed differences in the recombination landscape have also been influenced by other factors such as genomic structure. In order to fully understand the dynamics of recombination rate evolution, it will be necessary to obtain high-quality and fine-scale genetic maps across a wide range of species. As such maps become available, it will be possible to place the canine map into a proper evolutionary context, and thereby identify the factors that determine the forces that shape the distribution of recombination in the genome.

## Materials and Methods

DNA sequencing of 51 village dogs was performed using Illumina technology to 8-12 fold coverage, using 101 base-pair paired end reads (Supplementary Table 1 and Supplementary Table 2). Reads were aligned to the reference genome using *bwa* [31]. Variant calls were made using *GATK* [32], and phased using *BEAGLE* [33]. Extensive filters were applied to ensure that only high quality variants were used for the purposes of recombination rate estimation (see Supplementary Material for details). After filtering, recombination rates were estimated using the statistical package, *LDhat* [15].

For H3K4me3 ChIPseq experiments, spermatocytes of various stages were cell types were purified sedimentation velocity (STA-PUT) of collagenase digested single cell suspensions. Chromatin immunoprecipitation (ChIP) of H3K4me3 was performed using standard procedures. Libraries were prepared using Tru-Seq adaptors, with sequencing performed using 150bp paired-end reads from an Illumina HiSeq 2500 in Rapid Run mode. Reads were mapped to the canine reference genome, and H3K4me3 peaks called using *MACS* [34].

Detailed methods are available in the Supplementary Information.

## Data availability

Genetic maps and called hotspots are available for download from:
http://autonlab.einstein.yu.edu/dog_recomb/

## Acknowledgements

We would like to thank collaborators J. Calero, W. Johnson and R.K. Wayne for contributing wild canid DNA samples for analysis, and the efforts of M. Yee and the Stanford Sequencing Core. We also thank M. Castelhano, E. Corey and the Cornell DNA Bank and Medical Genetic Archive for DNA extraction and banking, and for countless students, veterinarians, collaborators and dog owners that contributed samples to the project.

## Figure Legends

Figure 1: The distribution of recombination. A) Genetic maps for chromosome 2 estimated using LD (red) and pedigrees [19] (blue). Other chromosomes are shown in Supplementary Figure 2. B) Genome-wide broad-scale recombination rates for each autosomal chromosome (red and orange). Rates were smoothed at the 5Mb scale. C) The concentration of recombination rate in dogs shows a more uniform distribution than human. Each dog chromosome is shown as a separate colored line, whereas estimates from HapMap [22] are shown in black. See also Supplementary Figure 5. D) Fine-scale recombination rates compared between dog (blue) and human (red) over a largely syntenic 60Mb region [35].

Figure 2: Recombination around genome features. A) Average recombination rates (on a logarithmic scale) for DNA repeats as a function of average GC content. Bubble size gives an indication of the number of repeats in a given family, with colors indicating higher-level repeat classes. Recombination rates for each repeat were estimated in a 5kb window centered on the repeat, thinned so that no two repeats were within 10kb. The insert shows (log scale) recombination rates around a selection of repeats. B) Recombination rates around TSS in dog (blue) shows an elevation that dwarfs the small elevation seen in human (red). C) Recombination

rates around TSS partitioned on the basis of distance to the nearest CpG island (as defined by the UCSC genome browser), ranging from genes with a CpG island overlapping the TSS (dark blue) to genes with no CpG island within 10kb (dark red).

Figure 3: Evidence of biased gene conversion in dog and fox genomes. The plots show the ratio of the number of AT->GC polymorphisms relative to the number of GT->AT polymorphisms around hotspots detected in dog that were localized to within 5kb. A) The (AT->GC)/(GC->AT) ratio around hotspots for SNPs discovered in dog (red) and fox (blue). 95% confidence intervals are shown as shaded areas, as assessed via bootstrap. B) The (AT->GC)/(GC->AT) ratio for polymorphisms originating along the dog lineage is stronger around more recombinogenic hotspots. The skew is shown around strong (red), intermediate (green), and weak (blue) hotspots, as defined by the peak rates described in the legend.

Figure 4: A) Recombination rates around H3K4me3 peaks, partitioned into peaks that contain a CpG island (red), and peaks without a CpG island (blue). B) The density of H3K4me3 marks around recombination hotspots partitioned into those hotspots that overlap a CpG island (red), and those that do not (blue). C) Recombination rates around CpG islands overlapping a H3K4me3 peak (red), and those not overlapping a H3K4me3 peak (blue).

## Tables Captions

Table 1: CpG-rich motifs are strongly enriched in canine recombination hotspots. Motifs were detected by comparing DNA sequences within 4,759 hotspots to the same number of coldspots, having matched for CpG content and SNP density. P-values are adjusted for multiple testing using the Bonferroni correction.

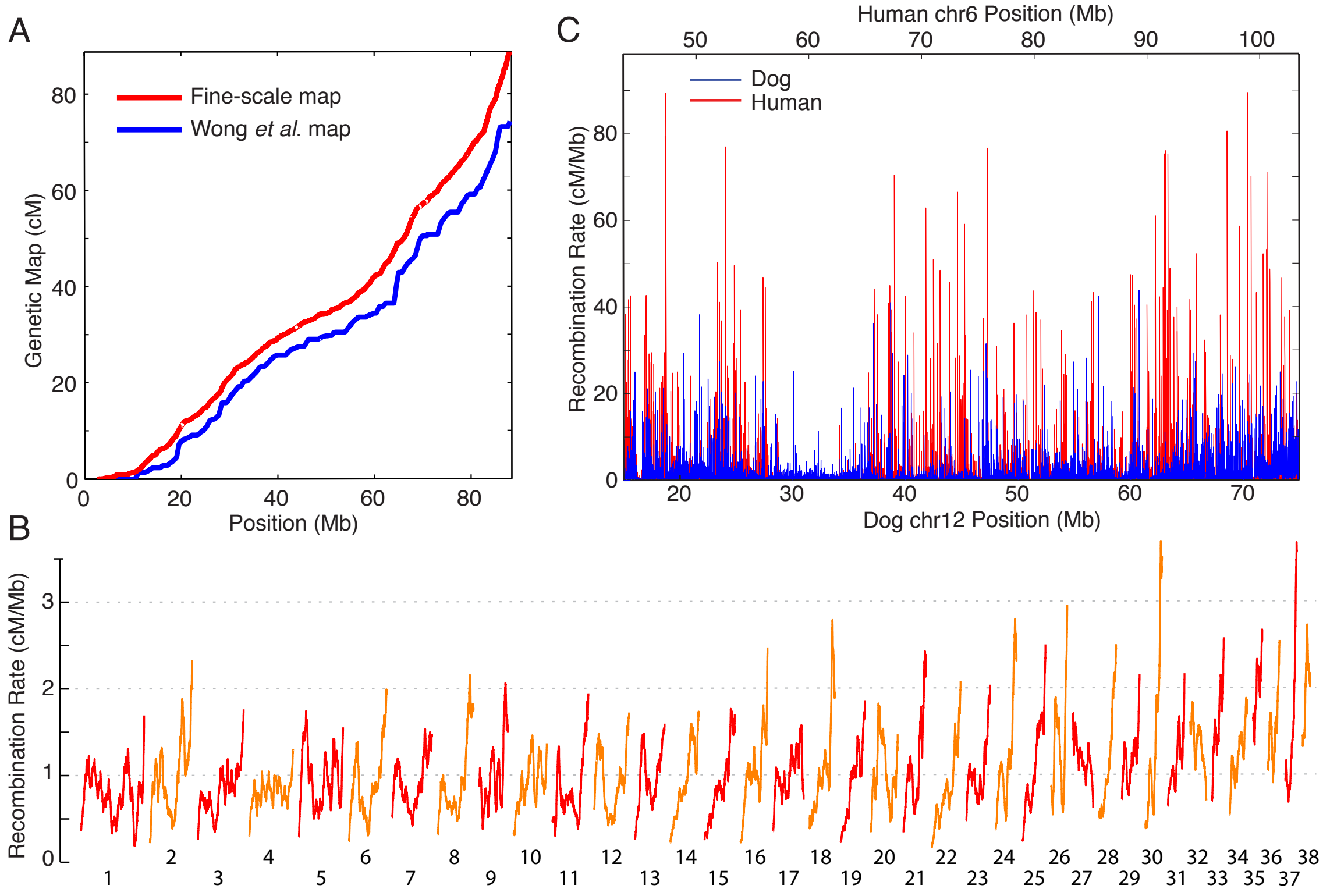

A
Fine-scale map
Wong et al. map
Genetic Map (cM)
Position (Mb)
B
Recombination Rate (cM/Mb)
C
Human chr6 Position (Mb)
Dog
Human
Recombination Rate (cM/Mb)
Dog chr12 Position (Mb)

Figure 1

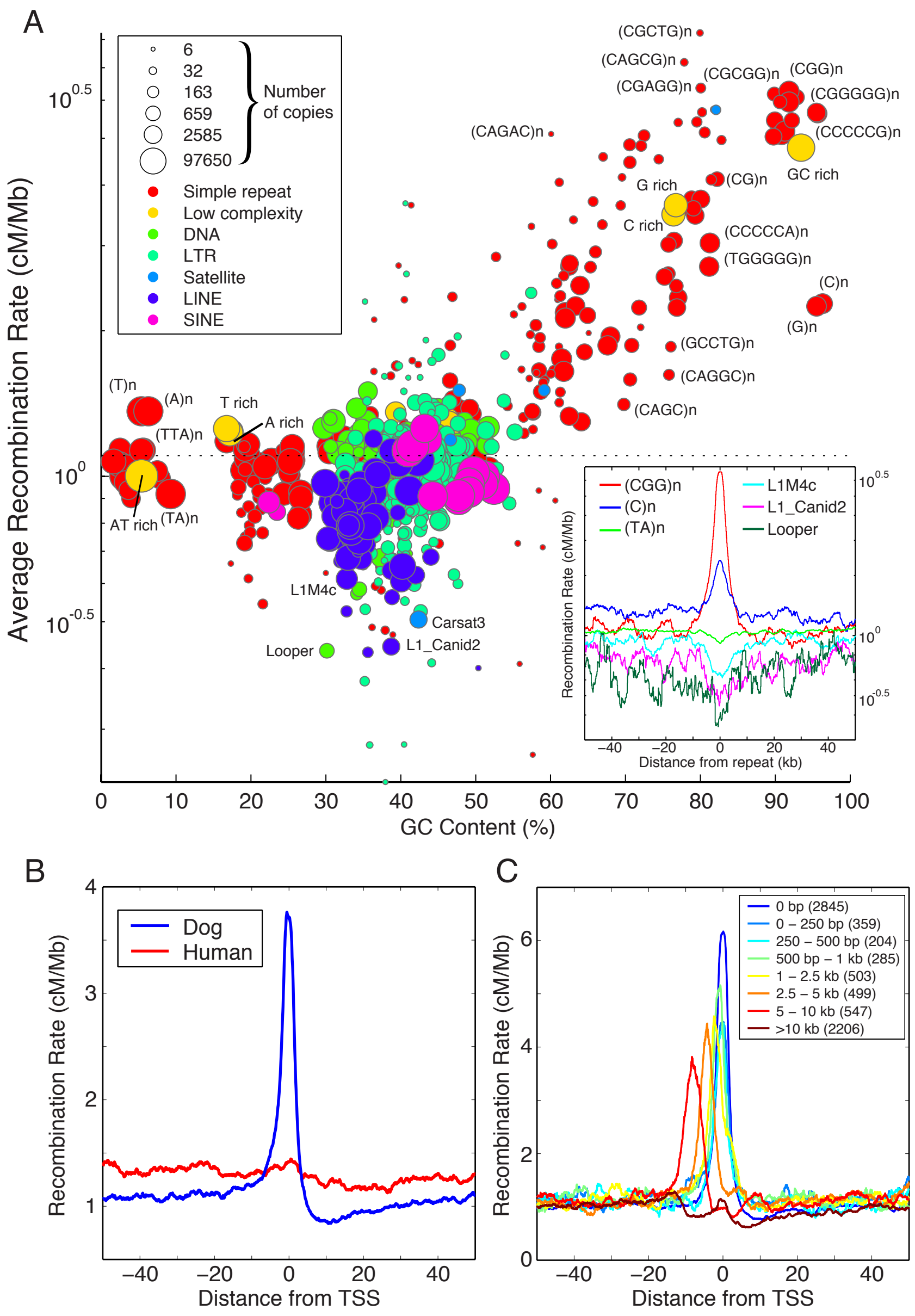

A
6
32
163
659
2585
97650
Number of copies
Simple repeat
Low complexity
DNA
LTR
Satellite
LINE
SINE
(CGCTG)n
(CAGCG)n
(CGAGG)n
(CGCGG)n
(CGG)n
(CGGGGG)n
(CAGAC)n
(CCCCCG)n
GC rich
G rich
(CG)n
C rich
(CCCCCA)n
(TGGGGG)n
(C)n
(G)n
(GCCTG)n
(CAGGC)n
(CAGC)n
(T)n
(A)n
T rich
A rich
(TTA)n
AT rich
(TA)n
L1M4c
Carsat3
L1_Canid2
Looper
Average Recombination Rate (cM/Mb)
GC Content (%)
Recombination Rate (cM/Mb)
Distance from repeat (kb)
B
Dog
Human
Recombination Rate (cM/Mb)
Distance from TSS
C
0 bp (2845)
0 – 250 bp (359)
250 – 500 bp (204)
500 bp – 1 kb (285)
1 – 2.5 kb (503)
2.5 – 5 kb (499)
5 – 10 kb (547)
>10 kb (2206)
Recombination Rate (cM/Mb)
Distance from TSS

Figure 2

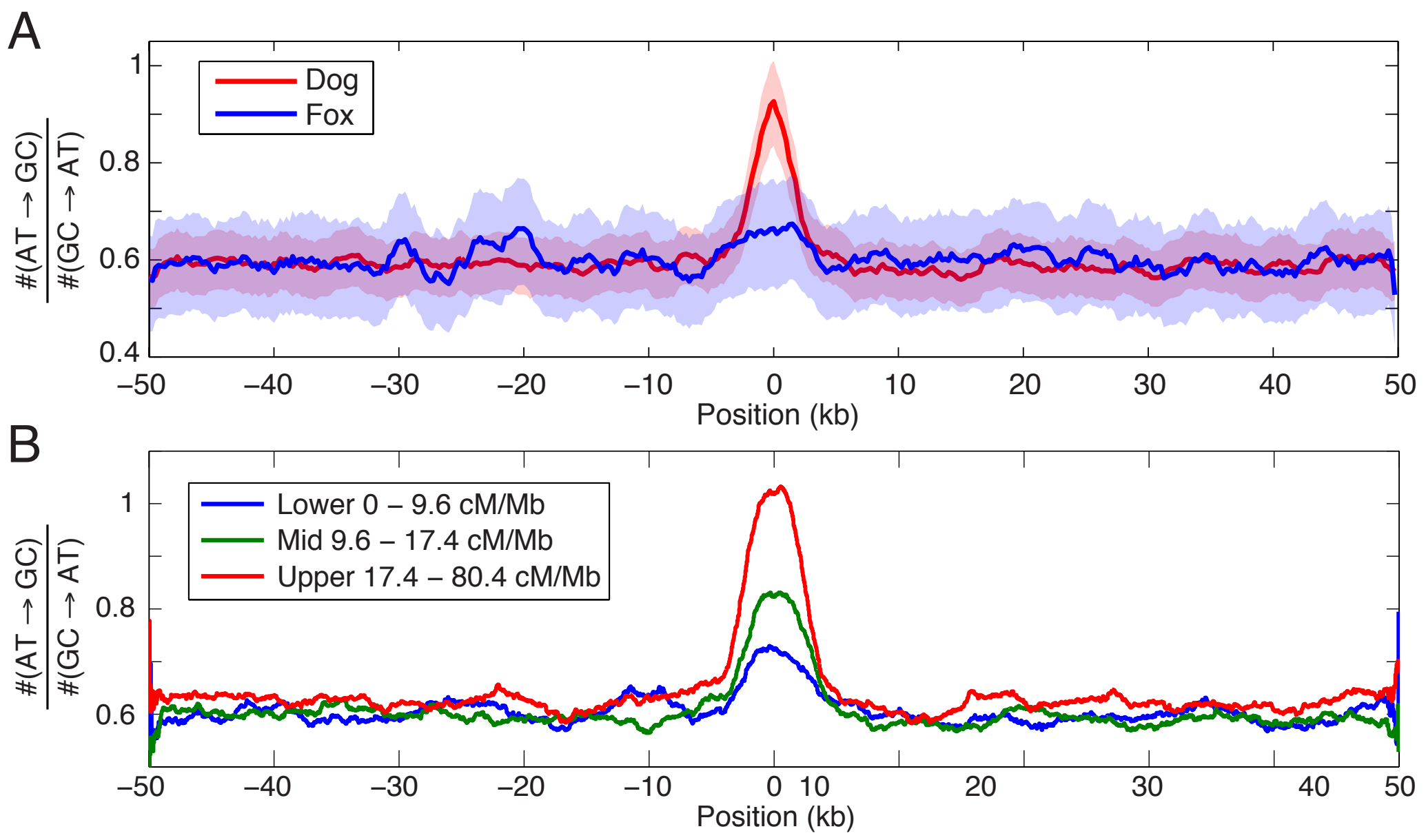
A
Dog
Fox
#(AT → GC) / #(GC → AT)
1
0.8
0.6
0.4
–50
–40
–30
–20
–10
0
10
20
30
40
50
Position (kb)
B
Lower 0 – 9.6 cM/Mb
Mid 9.6 – 17.4 cM/Mb
Upper 17.4 – 80.4 cM/Mb
#(AT → GC) / #(GC → AT)
1
0.8
0.6
–50
–40
–30
–20
–10
0
10
20
30
40
50
Position (kb)

Figure 3

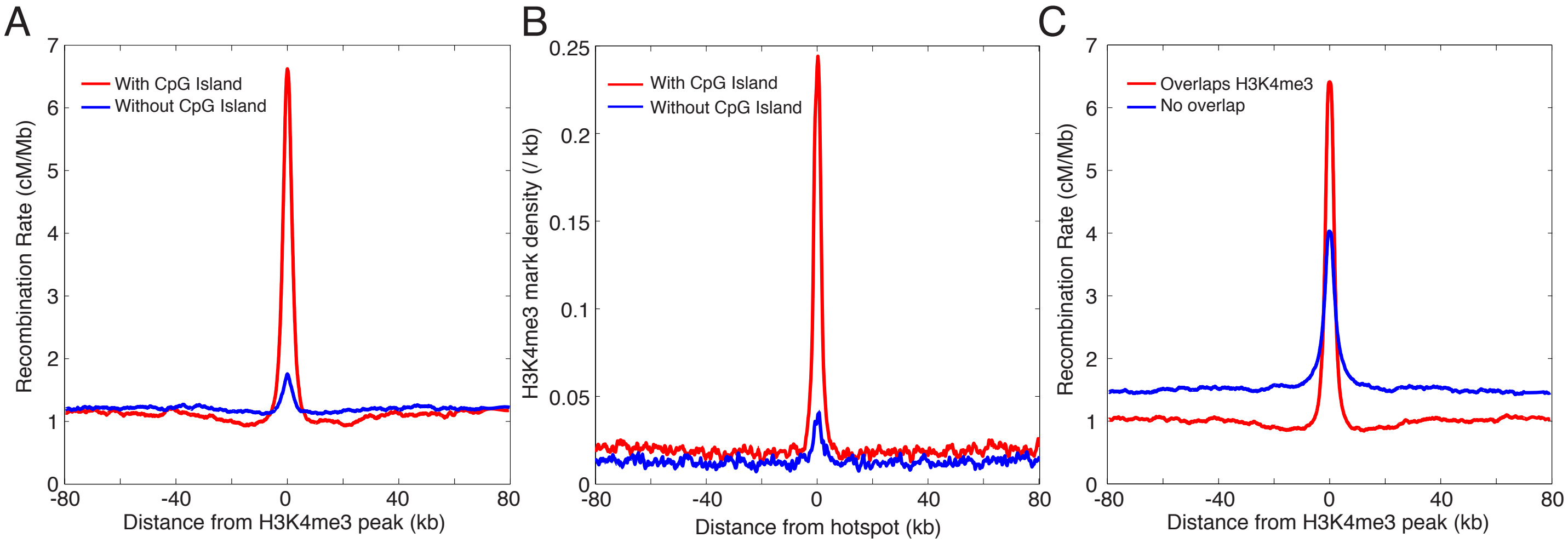

A
7
6
5
4
3
2
1
0
Recombination Rate (cM/Mb)
With CpG Island
Without CpG Island
-80
-40
0
40
80
Distance from H3K4me3 peak (kb)
B
0.25
0.2
0.15
0.1
0.05
0
H3K4me3 mark density (/ kb)
With CpG Island
Without CpG Island
-80
-40
0
40
80
Distance from hotspot (kb)
C
7
6
5
4
3
2
1
0
Recombination Rate (cM/Mb)
Overlaps H3K4me3
No overlap
-80
-40
0
40
80
Distance from H3K4me3 peak (kb)

Figure 4

**Table 1: Motifs enriched in hotspots.**

| Motif Length | Motif | Present in Hotspots | Present in Coldspots | Relative Enrichment | p-value (All Sequence) | p-value (Non-repeat Sequence) | p-value (Repeat Sequence) |
|---|---|---|---|---|---|---|---|
| 6 | CGCGCG | 20.7% | 13.5% | 1.54 | 1.24E-17 | 4.32E-14 | 2.35E-04 |
| 6 | CGCCGC | 43.4% | 35.8% | 1.21 | 1.48E-10 | 3.31E-15 | 2.84E-05 |
| 6 | CCGCGC | 38.3% | 31.2% | 1.23 | 9.58E-10 | 6.69E-11 | 3.15E-09 |
| 6 | CGCGGC | 39.5% | 33.5% | 1.18 | 2.31E-06 | 1.67E-07 | 1.37E-04 |
| 6 | CGGCCG | 30.8% | 25.8% | 1.19 | 1.42E-04 | 1.25E-04 | 1.59E-03 |
| 6 | CCGCCG | 42.3% | 36.9% | 1.15 | 1.48E-04 | 1.10E-07 | 1.09E-03 |
| 7 | CGCCGCG | 21.3% | 13.2% | 1.61 | 1.14E-21 | 1.51E-15 | 7.83E-08 |
| 7 | CCGCGCG | 16.4% | 9.8% | 1.67 | 1.94E-17 | 1.10E-10 | 3.42E-05 |
| 7 | CGCGCGC | 14.8% | 8.6% | 1.72 | 4.88E-17 | 8.27E-11 | 7.98E-03 |
| 7 | GCGCCGC | 27.4% | 19.3% | 1.42 | 7.59E-17 | 6.31E-14 | 8.86E-04 |
| 7 | CGCCCGC | 25.8% | 18.7% | 1.38 | 4.68E-13 | 1.12E-09 | 2.36E-06 |
| 7 | CCGCCGC | 29.1% | 21.9% | 1.33 | 4.57E-12 | 6.69E-12 | 3.82E-05 |
| 7 | CCGCGCC | 24.9% | 18.4% | 1.36 | 6.13E-11 | 2.71E-05 | 9.50E-15 |
| 7 | CGCGCCG | 17.3% | 11.7% | 1.48 | 7.11E-11 | 3.10E-03 | 1.49E-09 |
| 7 | CGCCGCC | 31.7% | 24.8% | 1.28 | 5.37E-10 | 2.20E-11 | 1.57E-03 |
| 7 | CCCGCGC | 25.7% | 19.4% | 1.33 | 9.90E-10 | 1.58E-04 | 2.35E-11 |
| 7 | CCGCGGC | 25.4% | 19.3% | 1.32 | 6.24E-09 | 5.13E-06 | 8.39E-07 |
| 7 | CCCGCGG | 26.4% | 20.8% | 1.27 | 1.07E-06 | 3.65E-04 | 3.84E-05 |
| 7 | GGCCGCC | 33.0% | 27.5% | 1.20 | 6.01E-05 | 8.96E-04 | 7.69E-03 |
| 7 | CGCGGGC | 24.9% | 20.0% | 1.25 | 9.52E-05 | 4.00E-03 | 2.38E-03 |
| 7 | CGGCGGC | 26.2% | 21.5% | 1.22 | 5.02E-04 | 9.50E-03 | 2.11E-03 |
| 8 | CGCCGCGC | 11.1% | 5.9% | 1.89 | 1.26E-15 | 7.96E-09 | 2.36E-03 |
| 8 | CCGCCGCG | 12.9% | 7.2% | 1.78 | 1.72E-15 | 8.54E-07 | 9.78E-06 |
| 8 | CCGCGGCG | 11.3% | 6.2% | 1.82 | 4.78E-14 | 4.39E-07 | 2.00E-04 |
| 8 | GCGCCCGC | 12.2% | 7.6% | 1.61 | 1.08E-09 | 2.64E-04 | 8.49E-03 |
| 8 | CGCGCCCC | 14.1% | 9.3% | 1.52 | 1.22E-08 | 8.16E-04 | 5.13E-04 |
| 8 | GCCGCCGC | 16.9% | 12.2% | 1.39 | 1.53E-06 | 5.18E-03 | 7.36E-03 |
| 8 | CCCGCCCC | 34.2% | 28.2% | 1.21 | 7.94E-06 | 3.10E-05 | 1.07E-04 |
| 8 | CCCCGCCC | 36.1% | 30.6% | 1.18 | 3.08E-04 | 1.55E-05 | 1.42E-03 |
| 9 | CGCCCCGCC | 10.5% | 5.6% | 1.89 | 6.11E-14 | 2.82E-07 | 3.42E-03 |
| 9 | CCCCGCCCC | 26.5% | 20.1% | 1.32 | 2.10E-08 | 4.60E-06 | 2.27E-03 |

# Supplementary Material

## Sample collection

Dogs were sampled between 2007 and 2010 at multiple locations throughout the world (Supplementary Table 1). We selected approximately 6 samples from various regions for sequencing, focusing mainly on 'free-ranging' village dogs which are generally more genetically diverse due to the absence of breed founder effects and artificial selection [17]. Dogs from China were selected for sequencing in an effort to maximize the geographical spread of the samples (1-2 for most provinces), and some of these are associated with breeds.

DNA was collected from 3-5 ml blood samples under Cornell IACUC #2005-0151, 2007-0076 and 2011-0061. DNA was extracted from a captive female culpeo fox (*Lycalopex culpaeus culpaeus*) from the Cusco region of Peru using a similar procedure.

## Sequencing

Paired end libraries were prepared from genomic DNA using the standard Illumina protocols. Briefly, genomic DNA was sheared in a Covaris E210. Sheared DNA was end-repaired, A-tailed, gel-purified and ligated with Illumina's paired end adapters. Libraries were amplified for 8 cycles. Libraries were run for 2 x 101 bp reads on one lane each on a HiSeq 2000. This procedure resulted in between 8X and 12X coverage for each sample.

## Mapping and variant calling

The CanFam3.1/canFam3 reference assembly was downloaded from the Broad FTP site [36] in October 2010 and converted to FASTA format. We note that this reference differs from the final CanFam3.1/canFam3 reference released by UCSC due to differences in choices regarding padding between contigs. Sequence reads from 57 dogs were mapped to reference using *bwa* [31]. PCR duplicates were excluded using *Picard* [37], and empirical quality recalibration and realignment around candidate indels was performed using *GATK* [32]. SNP genotype likelihoods were generated by *GATK* and passed to *BEAGLE* [33] for genotype calling and phasing.

The Andean fox sequence was processed in a similar manner, with reads also aligned to the canine reference genome. *GATK* was also used to generate genotype calls at dog variant sites, yielding a 95% call rate with approximately 1% heterozygous genotype calls. These variant calls were used to polarize the ancestral allele for polymorphisms discovered in dog, with 3,296,170 sites being assigned an ancestral allele with high confidence. In addition, *GATK* was also used to perform

SNP discovery in the fox, allowing discovery of 1,287,138 heterozygote polymorphisms in the fox with a SNP quality > 100 and a genotype quality >= 99.

## Calculation of variant detection power and genotype accuracy

A total of 48 of the 51 sequenced dogs were genotyped on Illumina CanineHD microarrays [18]. After filter out sites with low call rates (<90%), low duplicate concordance rates across genotyping platforms, or sites at positions which did not uniquely map from canFam2 to canFam3 with liftOver, we estimate that we have >98% power to detect variants with minor allele frequency of 5%, and a nearly 95% power to detect rarer variants. For these variants detected by both genotyping and sequencing approaches, the overall concordance of the genotype calls is 99.1%.

## Construction of a genetic map

Prior to estimating the genetic map, we conducted extensive filtering of the variants called in the filtering. In part, this was because past experience has indicated that estimating recombination rates from patterns of LD can be moderately sensitive false-positive variant calls [10]. Specifically, we filtered variants to exclude all indels, and non-biallelic variants. In addition we filtered SNPs with unusual depth of coverage (average coverage across individuals <6X or >12X), or highly diverged from Hardy Weinberg Equilibrium (the number of heterozygote genotypes was < 60% or > 100% than the expectation under HWE). In order to focus on only very high quality variants, we excluded any variant in a homopolymer of length > 3.

The resulting callset consisted of 3,560,506 polymorphic SNPs on the autosomes, 145,431 SNPs on the non-pseudoautosomal X chromosome, and 18,285 SNPs in the pseudoautosome.

The genetic map was estimated using the *LDhat* [15,38] software package, which uses a coalescent-based model to infer historical recombination rates from population genetic data. To estimate the genetic map, we followed a procedure similar to that used previously [10,14,15,22]. Specifically, we split the dataset into 4000 SNP windows with a 200 SNP overlap between windows. Recombination rates were estimated by *LDhat* for each window independently using a block penalty of 5 with 60 million MCMC iterations. A sample was taken from the chain every 40,000 iterations, and the first 500 samples (corresponding to 20 million iterations) were discarded as burn-in.

In addition, preliminary analysis revealed a strong association with CpG-rich regions of the genome. Therefore, all recombination analyses have been repeated having excluded all putative CpG sites (i.e. either SNP allele would result in a CpG site), as well as any SNP within 1kb of a putative CpG island, from the call set to ensure that the resulting observations were robust to any CpG-associated artifacts. We also investigated stronger filters that those described above, including filtering out sites with low minor allele frequencies or mapping to repetitive elements in the genome, but found them to have a minimal qualitative impact on the final estimates. Likewise, we conducted investigations regarding which dogs to include in the dataset when generating the genetic map. We found that using population subsets of dogs did not qualitatively alter patterns observed in final estimates, and hence we chose to include dogs that are not strictly village dogs as long as their inbreeding was not too severe (F > 50%). Three purebred dogs (a basenji, a Mongolia shepherd, and Perdiguero) were excluded as were three New Guinea Singing dogs either due to high inbreeding or due to being outliers in PC1 and PC2 when the genotype data was analyzed using principal component analysis. We have replicated the results described in this paper using more conservative call sets with more selective dog and SNP selection procedures, and have found them to be robust.

*LDhat* provides estimates of the population recombination rate, $\rho = 4N_e r$. As *LDhat* estimates recombination rates from patterns of linkage disequilibrium (LD), artifactual breakdowns in LD can result in artificially large, and biologically implausible, estimates of recombination [10]. To this end, we set the recombination rate to zero within a region if the estimate of $4N_e r$ between a pair of adjacent SNPs was greater than 100, or if there was a gap in the reference of greater than 100kb in size. Whenever such a region was identified, the recombination rates for the surrounding 100 SNPs (50 SNPs in both directions) were also zeroed out, as these estimates are also likely to unreliable. In total, these filters zeroed out recombination rates in 0.12% of SNP intervals (arising from 2 gaps > 100kb, and 40 distinct regions with a SNP interval $4N_e r$ estimates > 100).

In order to convert the population estimate into a per-generation recombination rate estimate (measured in units of cM/Mb), it is necessary to obtain an estimate of the effective population size, $N_e$. In order to do this, we used a robust linear regression (without intercept) between the *LDhat* estimates and the experimental estimates obtained from Wong *et al*. [19] in 5 Mb bins. The gradient of the fitted line can be used to obtain an estimate of $N_e$. The use of robust regression ensures that local deviations in the correlation between human and dog do not overly influence the $N_e$ estimate. Using this method, we estimated of $N_e$ as 31034.71, which we used to scale the *LDhat* estimates accordingly.

## Calling hotspots

Recombination hotspots in the dog genome were called in a similar manner as used previously for chimps [10]. Briefly, to assess if observed peaks in the estimated recombination rate estimates represent significant variation over and above the noise in the estimator we used coalescent simulations to assess the significance of recombination peaks detected in the data. The method detects regions of localized elevated recombination rate by comparing a model in which the recombination rate within a small region at the center of a window is equal to the surrounding background recombination rate (the null model) to a model in which the recombination rate is allowed to take any value (the alternative model). The algorithm uses coalescent simulations to determine the null distribution of a test statistic in the absence of a recombination hotspot, and compares this distribution to the test statistic obtained from the true data. A hotspot is called if the true data test statistic lies in the extreme tail of the empirical null distribution. The test statistic in this case is the composite likelihood used by *LDhat* [15,38].

The hotspot-detection method processes the data in windows of 100kb, testing the central 3kb for the presence of a recombination hotspot. The window is moved 1kb at each iteration. For each window, a maximum likelihood constant background recombination rate is calculated across the 100kb window. Subsequently, a maximum likelihood estimate of the recombination rate was obtained allowing the central 3kb to take vary from the background. The test statistic is taken as (composite) likelihood ratio between the two models. In order to obtain the null distribution of the test statistic, we performed coalescent simulations using a constant recombination rate drawn from an exponential distribution with mean equal to the background rate measured in the real data. We performed 5,000 simulations for each putative hotspot, although the simulations were cut short if there is no evidence of significance after 50 simulations ($p > 0.01$). Simulations were conducted using the neutral, equilibrium coalescent with recombination and assuming the infinite sites model. The number of mutations was conditioned to match the number of SNPs in the real data, and mutations were placed at SNP locations.

Having tested all 3kb windows in the genome, hotspots were called using the following steps. First, all 3kb windows with $p<0.01$ were selected. Adjacent or overlapping windows were merged to provide a list of putative hotspot regions. After merging, we discarded regions that did not contain at least one window with a p-value < 0.001. At this stage, we called 7,677 putative hotspots with mean width of 21,954 bp (10,000 bootstrap 95% C.I. 21,696 – 22,225 bp).

To further localize the called hotspots, we compared the hotspot regions to the recombination rate estimates obtained from *LDhat*. Specifically, the peak rate was taken as the maximum rate within the significant region, and the boundaries of the peak were taken as the Full Width at Half Maximum (FWHM) if this was smaller than the original significant region boundaries. This procedure resulted in a mean hotspot width of 8,259 bp (10,000 bootstrap 95% C.I. 8,066 – 8,467 bp).

For each hotspot, we considered the GC, CpG, and N (missing) base composition of the reference sequence contained with the hotspot. We discarded hotspots if more than 1% of the reference sequence was missing within the full width region, leaving 5,467 hotspots. The mean FWHM width of these hotspots was 7,540 bp, with 4,255 hotspots (78%) localized to within 10kb, and 2,927 (54%) localized to within 5kb.

## Identifying hotspot-associated motifs

For each hotspot, we attempted to identify a nearby region showing no evidence for recombination rate elevation (a 'coldspot'), matched for GC content and SNP density. We first identified all 3kb windows of the genome for which there was no evidence of a recombination hotspot ($p>0.05$). We removed all putative coldspots containing more than 1% missing sequence.

To match each hotspot with a corresponding coldspot, we first estimated the GC and CpG content of each hotspot in a 3kb window centered on the hotspot peak center. For each hotspot, we then identified all coldspots on the same chromosome with a GC content within 0.5% of the hotspot GC content and CpG content within 0.1%. If more than one coldspot met these criteria, the coldspot matching the hotspot closest in terms of SNP density was chosen, as measured within a 20kb windows centered on the hotspots and coldspots. Using this method, we were able to identify a matched coldspot for 4,759 hotspots.

We extracted the DNA sequences associated with the 3kb windows around the center of the hotspots and coldspots. We tested all motifs with lengths between 3 and 10 base pairs inclusive. For each motif, we calculated the number of hotspots and coldspots having at least one copy of the motif, and calculated the significance of the difference using Fisher's Exact Test. Reported p-values were Bonferroni corrected within in each motif length class. We repeated the procedure twice: once having masked out repeat sequence (as defined by RepeatMasker), and once having masked out repeat sequence. The motifs reported in Table 1 show significance of $p < 0.01$ after Bonferroni correction in both repeat and non-repeat sequence.

## Comparison of the distribution of recombination in dogs to human

Our recombination rate estimates suggest that dogs have a more uniform distribution through the genome than that observed in human (Supplementary Figure 5A). However, we note that the estimated effective population size in dogs is higher than has been reported in human. In order to investigate if the observed distribution in dogs could be driven by this difference in effective population size, we conducted a simulation study. Using coalescent simulations, we simulated data under 3 different effective population sizes; 10,000, 20,000, and 30,000. The simulated dataset consisted of a 250kb region with a central 0.2cM hotspot, and a background rate of 0.0125 cM / Mb. Simulations for each effective population size were repeated 250 times. The results are shown in Supplementary Figure 5B. As is clear, the effective population size can have a significant effect on the distribution of recombination, and hence it cannot be concluded that observed differences in the distribution of recombination between dog and human represent true differences in the underlying distribution of recombination.

## H3K4me3 Chromatin Immunoprecipitation and sequencing (ChIP-seq)

Canine testes were obtained from routine neutering procedures from either the Cornell University Hospital for Animals, or the Tompkins County SPCA, and were stored for less than 12 hours in 1x PBS buffer in 4°C. The surrounding capsule and epididymis were removed and the testis cut up into 1 cm pieces with a scalpel blade, which were then minced, collagenase treated and added to Krebs buffer. The cell types were purified using gravity sedimentation, as a single cell suspension, according to the protocol described previously [39]. Briefly, the testis slurry was digested in collagenase and then trypsin to generate a single cell suspension, which was then run through a gravity sedimentation chamber at 4°C for several hours. The resulting sediment was fractionated and fractions examined for cellular content using a light microscope. The spermatocytes of specific prophase I stages were determined by their sizes and morphology. Cells were cross-linked in 1% formaldehyde and flash frozen.

Chromatin immunoprecipitation (ChIP) of histone H3 lysine 4 trimethylation (H3K4me3, using antibody ab8580 from Abcam) was performed using chromatin from approximately 2 million canine spermatogenic cells (pachytene and leptotene/zygotene spermatocytes were handled separately). ChIP was carried out according to the Myers laboratory protocol [40]. Prior to the incubation of chromatin with the antibody-coupled beads, one-tenth of the chromatin sample was removed to use as input for ChIP-seq. During the overnight incubation at 65°C to reverse cross-linking, 5 M NaCl and Proteinase K were added to the reaction. ChIP was validated through qPCR, with primers as described in Supplementary Table 6.

Illumina ChIPseq libraries were prepared using Tru-Seq adaptors. Sequencing was performed on input and ChIP samples for both cell types using 150 bp paired-end sequencing on a lane of an Illumina HiSeq 2500 sequencer in Rapid Run Mode.

Reads were subsequently mapped to the canine reference genome using *bwa* [31] (0.6.2-r126). H3K4me3 peaks were called using *MACS* [34] assuming a reference genome size of 2.2e9 and a bandwidth of 150.

## Polarization of dog and fox SNPs

Biased gene conversion is a proposed mechanism by which transmission of G/C alleles can occur preferentially over A/T alleles in the vicinity of double strand breaks associated with recombination (Supplementary Figure 13A). To observe the effect of biased gene conversion, it was necessary to polarize SNP polymorphisms by the ancestral allele. This is complicated by the lack of reference genome for the fox. Therefore to assign ancestral alleles, we used an *ad hoc* approach.

To assign ancestral alleles in dog, we called genotypes in the fox at all sites called in the dog. We assigned the reference allele as ancestral if the fox was called as homozygous for the reference allele, and assigned the alternative allele as ancestral if the fox was called as homozygous for the (dog) alternative allele. We required at least 5 reads and a genotype quality of at least 10 in the fox, and did not assign ancestral alleles at other sites.

For the converse process of assigning an ancestral allele for fox SNP polymorphisms, we extracted all positions in which the fox was heterozygotic with one of the alleles being the dog reference base. In this case, we took the ancestral allele as the dog reference base. We did not assign ancestral alleles when the fox was homozygotic for either the reference or alternative alleles.

Having defined ancestral alleles, we were able to investigate patterns of biased gene conversion. We define the 'skew' of AT->GC mutations relative to GC->AT mutations as the ratio of the number of observed mutations in each class.

$$\text{Skew} = \frac{\#(AT \rightarrow GC)}{\#(GC \rightarrow AT)}$$

To investigate the properties of the 'skew' statistic, we performed a simulation study to investigate the effect of biased gene conversion. We used the SFS_CODE software package [41], which provides a highly flexible forward simulator that can model the effects of biased gene conversion. We simulated polymorphism data in 50 individuals within 5000 10kb regions, each containing a central hotspot with a 1kb with and a $4N_er$ across the hotspot of 17. We repeated the simulations both with and without biased gene conversion. In the biased gene conversion case, parameters were selected to match estimates regarding BGC obtained in the literature, such that 90% of recombination events result in a gene conversion with a mean tract length of 150bp. The allele bias was taken as 0.83, matching experimental sperm typing estimates [8]. We then estimated the #AT->GC / #GC->AT ratio averaged across the hotspots, using only sites segregating within the samples.

The results are shown in Supplementary Figure 13B. In the absence of biased gene conversion, there is no change in the skew statistic in the vicinity of the hotspot. However, in the presence of biased gene conversion, a clear peak can be seen, qualitatively similar to the pattern observed around canine hotspots.

## Genome annotations

Gene annotations and were downloaded from Ensemble in canFam2 coordinates. DNA repeat and CpG island were downloaded from the UCSC genome browser, also in canFam2 coordinates. These annotations were lifted to the coordinates of our reference build using the UCSC liftover tool (having generated the appropriate 'chain' files). Annotations that uniquely lifted over to our reference were kept (30,173 genes).

## Sanger Sequencing of *PRDM9* in foxes

We sequenced the zinc-finger encoding exon 7 of *PRDM9* in an Island Fox, *Urocyon littoralis,* and a Culpeo (or Andean Fox), *Lycalopex culpaeus*. PCR sequencing was performed using the same procedure described in Axelsson *et al*. [13].

## Supplementary Figure Legends

Supplementary Figure 1: Decay of LD for village dogs compared to breed dogs. Genotype LD was calculated between the village dogs (red hued lines), breed dogs (blue hued lines), and wolves (green line). The genotype data for wolves and breed dogs was taken from Axelsson *et al*. [13]. For the purposes of comparison, the sample size in each group was N=6, and only sites present on the Illumina SNP microarray used by the Axelsson study were used. LD was calculated using the PLINK [42] for sites with MAF > 10%.

Supplementary Figure 2: Genetic maps for all autosomes estimated using LD (red) and pedigrees [19] (blue).

Supplementary Figure 3: Correlation at the 5Mb scale between LD-based recombination rate estimates and those obtained from with Wong *et al [19]*. Each chromosome is indicated with a different marker.

Supplementary Figure 4: A) Correlation between genetic map length and chromosome length for the LD-based map (black) and the linkage (red). Lines indicate least-squares lines of best fit. B) Shorter chromosomes tend to have higher recombination rates in both maps. Colors as for (A). Lines indicate a line of best fit estimated for a cubic function.

Supplementary Figure 5: A) The distribution of recombination for each dog chromosome, compared to the human estimates from HapMap. B) Conclusions regarding differences between the human and dog distributions should be treated with caution due to sensitivity in the estimator to the effective population size. This figure shows the distribution of recombination for data simulated with three different effective population sizes (Ne).

Supplementary Figure 6: Pearson correlation at various scales between recombination rate estimates from this study, Axelsson *et al*. [13], and Wong *et al*. [19].

Supplementary Figure 7: A) Recombination rate estimates from the Axelsson *et al*. study around hotspots identified by this study. Estimates are shown for all hotspots (blue), those called by both studies (green), and those not called by both studies (red). B) Rate estimates from this study around hotspots identified in the Axelsson *et al*. study. Colors as for (A).

Supplementary Figure 8: The association between recombination and GC and CpG content. A) GC content verse normalized recombination rate at 1kb, 10kb, and 100kb scales. Recombination rates are shown for dog (red) and human (blue), having normalized the rates in each species by subtracting off the mean and dividing by the standard deviation. B) As for (A), but showing CpG content verses normalized recombination rate. C) Pearson correlation of GC content with (un-normalized) recombination rate, as a function of scale. D) As for (C) but showing correlation between CpG content and recombination rate at various scales.

Supplementary Figure 9: Recombination in each region of a gene. Shown are recombination estimates 50kb Upstream of the transcription start site (TSS), within the 1st Exon, 1st Intron, Middle Exon, Last Intron, Last Exon, and 50kb Downstream from the transcription end site (TES). Estimates within each intron and exon were estimated across 50 equally spaced bins, allowing a percentage to express the distance through the intron / exon.

Supplementary Figure 10: Recombination rates around CpG islands partitioned on the basis of distance to the nearest transcription start site.

Supplementary Figure 11: A) No evidence of biased gene conversion around recombination coldspots. The plots show the ratio of the number of AT->GC mutations relative to the number of GT->AT mutations for polymorphisms identified in dog (red) and fox (blue). B) The same ratio around hotspots and coldspots in dogs, having excluded all polymorphisms for which either allele would create CpG dinucleotide.

Supplementary Figure 12: Recombination rates for H3K4me3 peaks overlapping CpG islands (A) and not overlapping CpG islands (B). The three lines represent peaks shared between pachytene and L/Z (blue), those unique to pachytene (green), and those unique to L/Z (red).

Supplementary Figure 13: A) Cartoon representation of a possible biased gene conversion mechanism. A double-strand break in the vicinity of a G/A polymorphism results in the partial loss of the A/T base pair on the 2nd chromosome. The subsequent repair following strand invasion results in mis-pairing between the G and T nucleotides. Biased gene conversion results from the biased repair of this mismatch, which favors the G allele over the T allele. B) Behavior of the 'skew' statistic in 5000 simulated hotspots, shown for simulations including biased gene conversion (red), and without biased gene conversion (blue).

## Supplementary Tables

Supplementary Table 1: Summary of dogs used to construct genetic map.

Supplementary Table 2: Summary of coverage in each dog.

Supplementary Table 3: Regression of recombination with GC content, CpG content and gene density at various scales.

Supplementary Table 4: *PRDM9* DNA sequences in dogs and foxes. This table shows the ~651 bp of DNA sequence from exon 7 of *PRDM9* in 24 species or breeds. Sequences in black were taken from Axelsson *et al*. [13]. whereas the sequences from this study are highlighted in red. Sequences were aligned in MEGA [43] version

5.05, specifically using the clustalW algorithm with standard parameters. Frameshift mutations are visible at positions 40 and 533 for the Andean Fox (Culpeo) respectively, with a stop codon at starting at position 130. The island fox has a stop codon starting at position 607.

Supplementary Table 5: PRDM9 protein sequences in dogs and foxes. Data is as for Supplementary Table 4, but translated to amino acid sequence. The frameshift mutations in the Andean Fox (Culpeo) can be seen at positions 14 and 178, with the stop codon at position 44. The Island Fox stop codon can be seen at position 203.

Supplementary Table 6: Primers used for qPCR validation of ChIP.

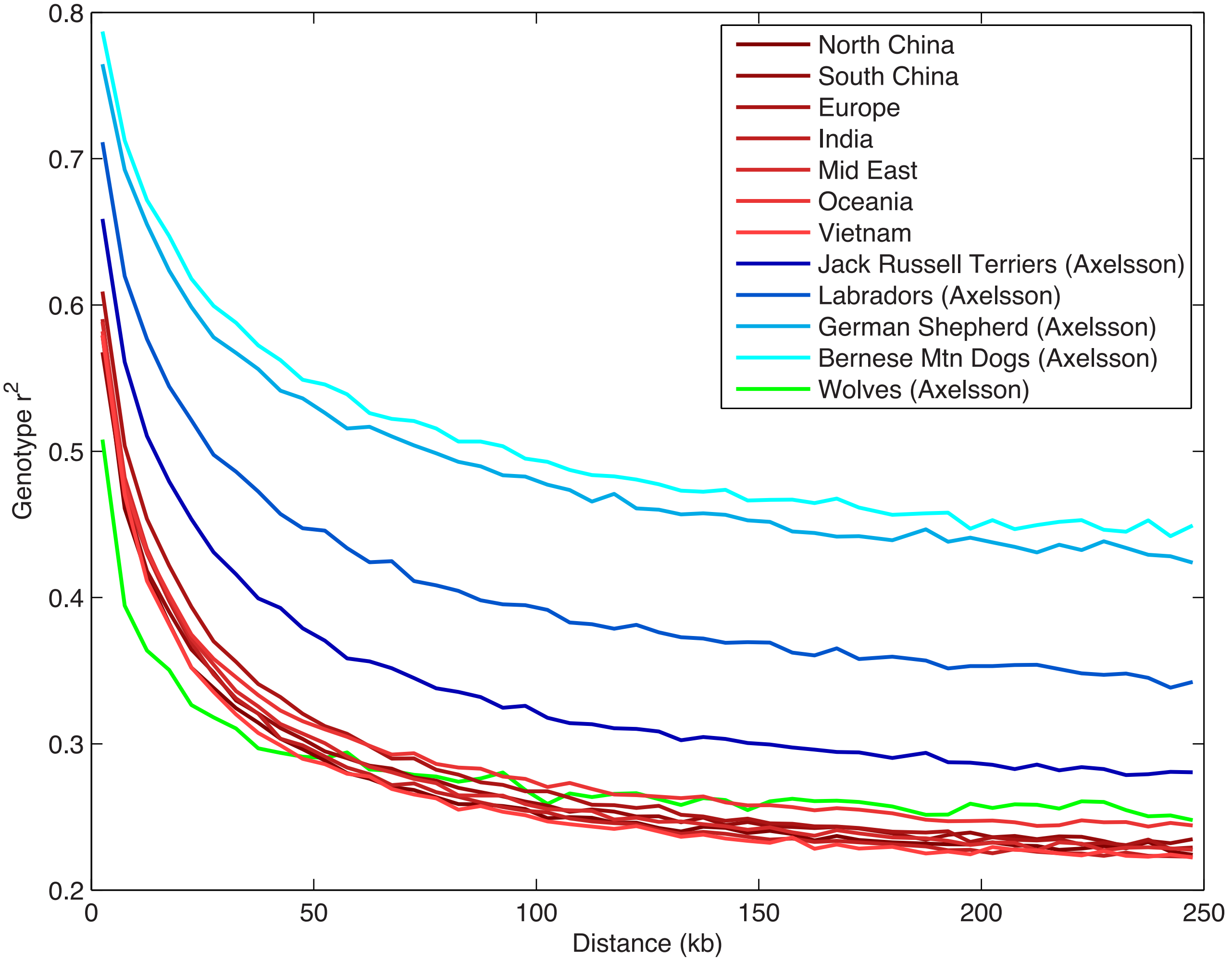

Supplementary Figure 1

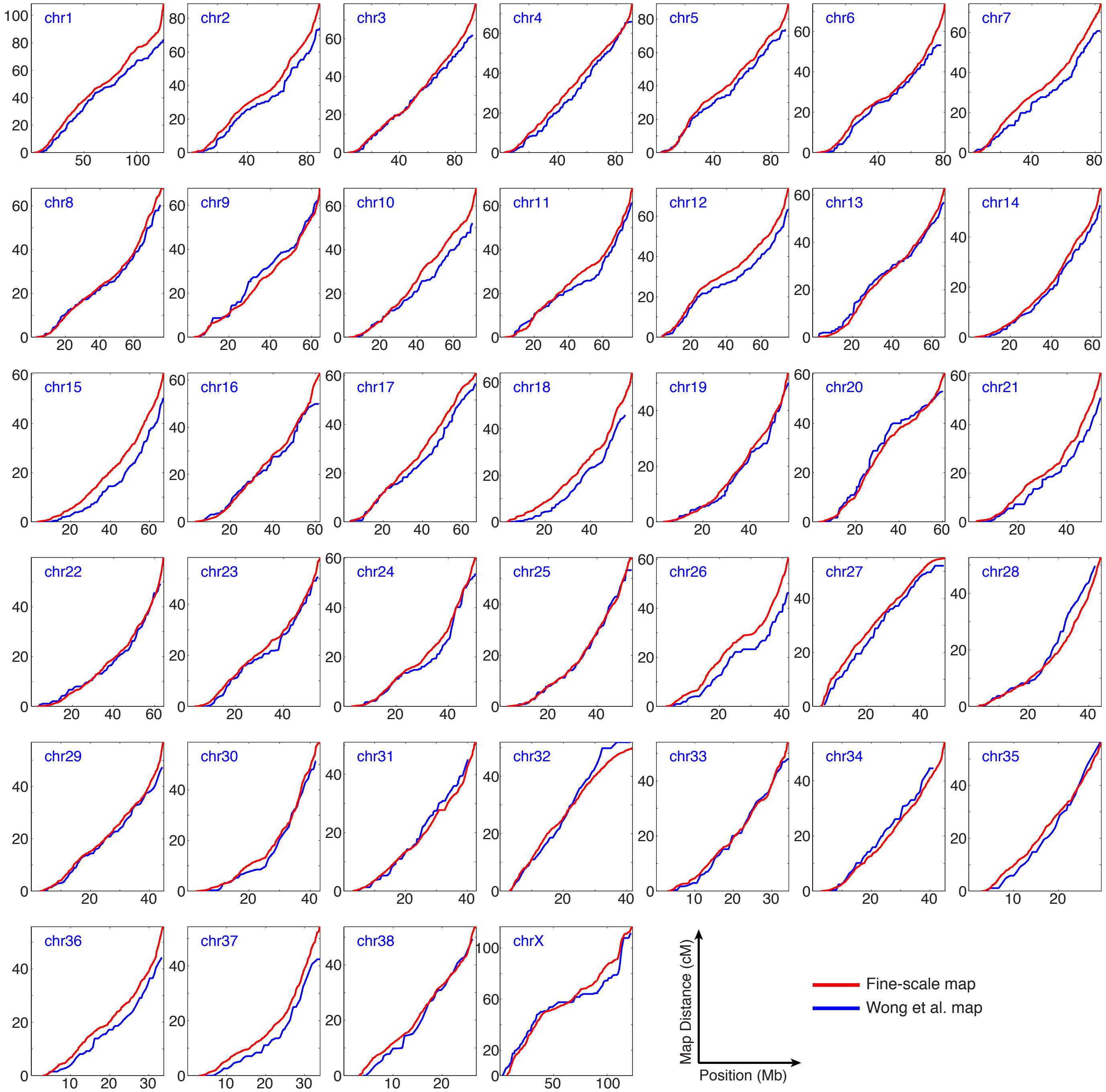

Supplementary Figure 2

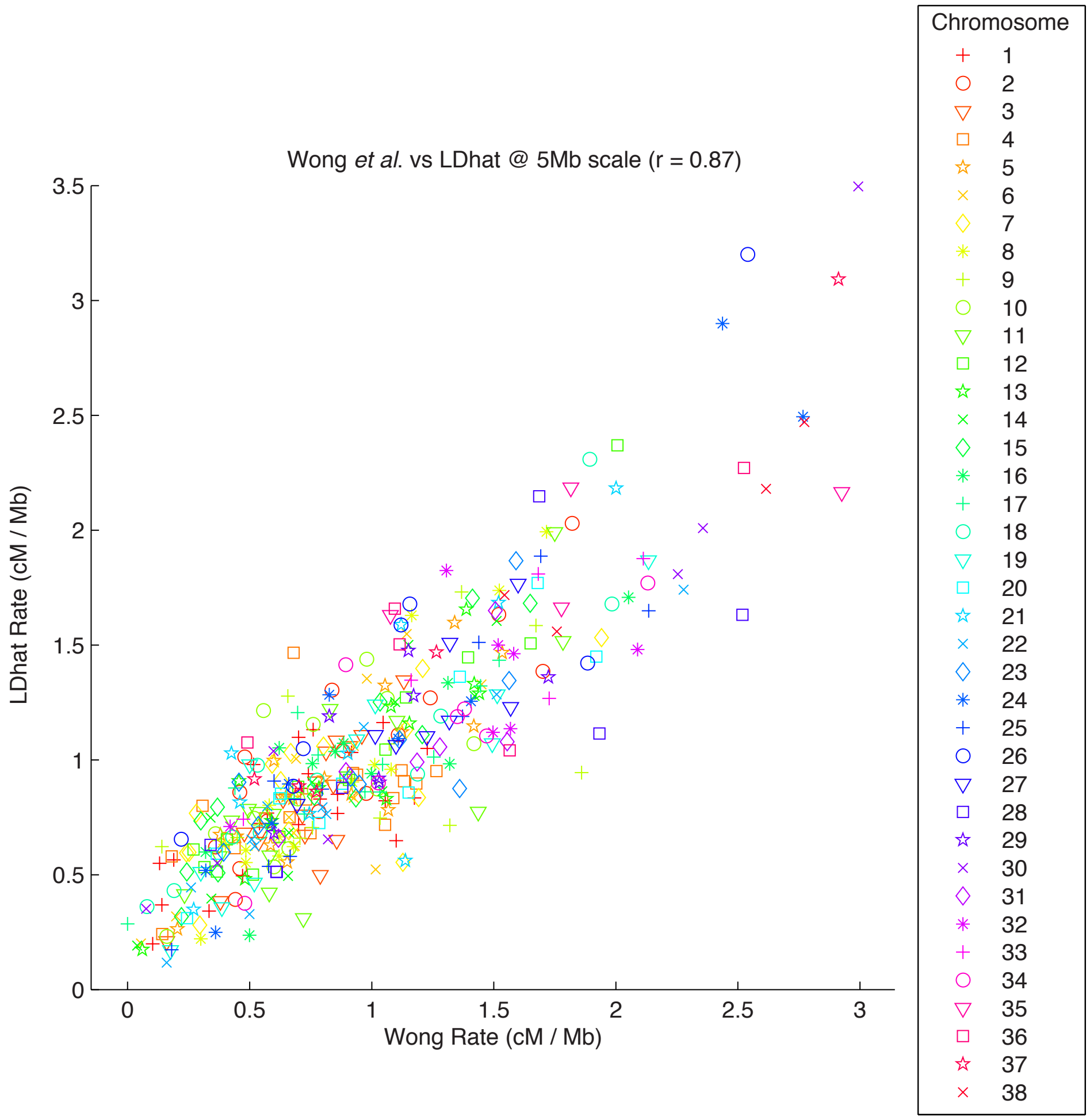

Supplementary Figure 3

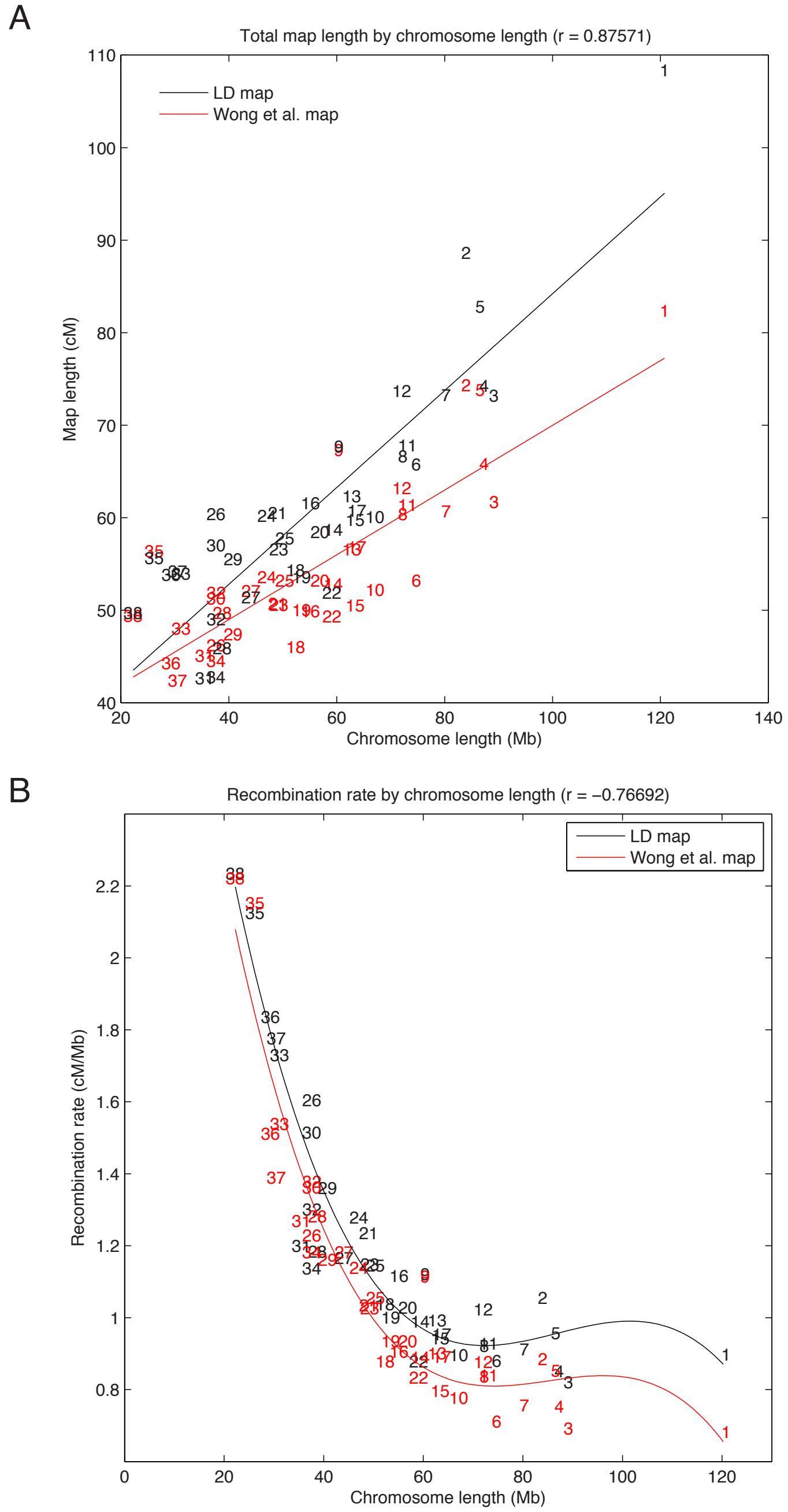

Supplementary Figure 4

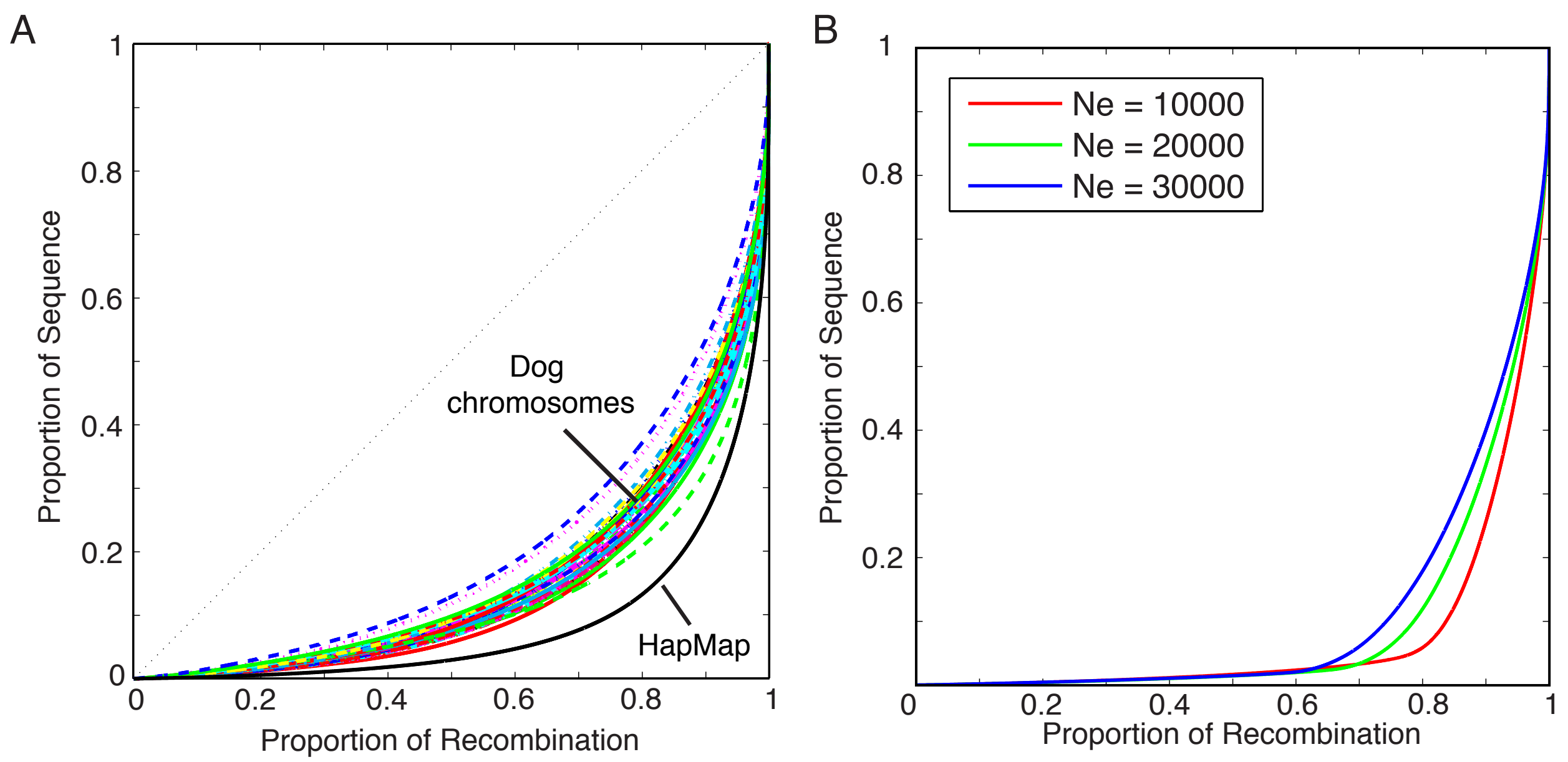

Supplementary Figure 5

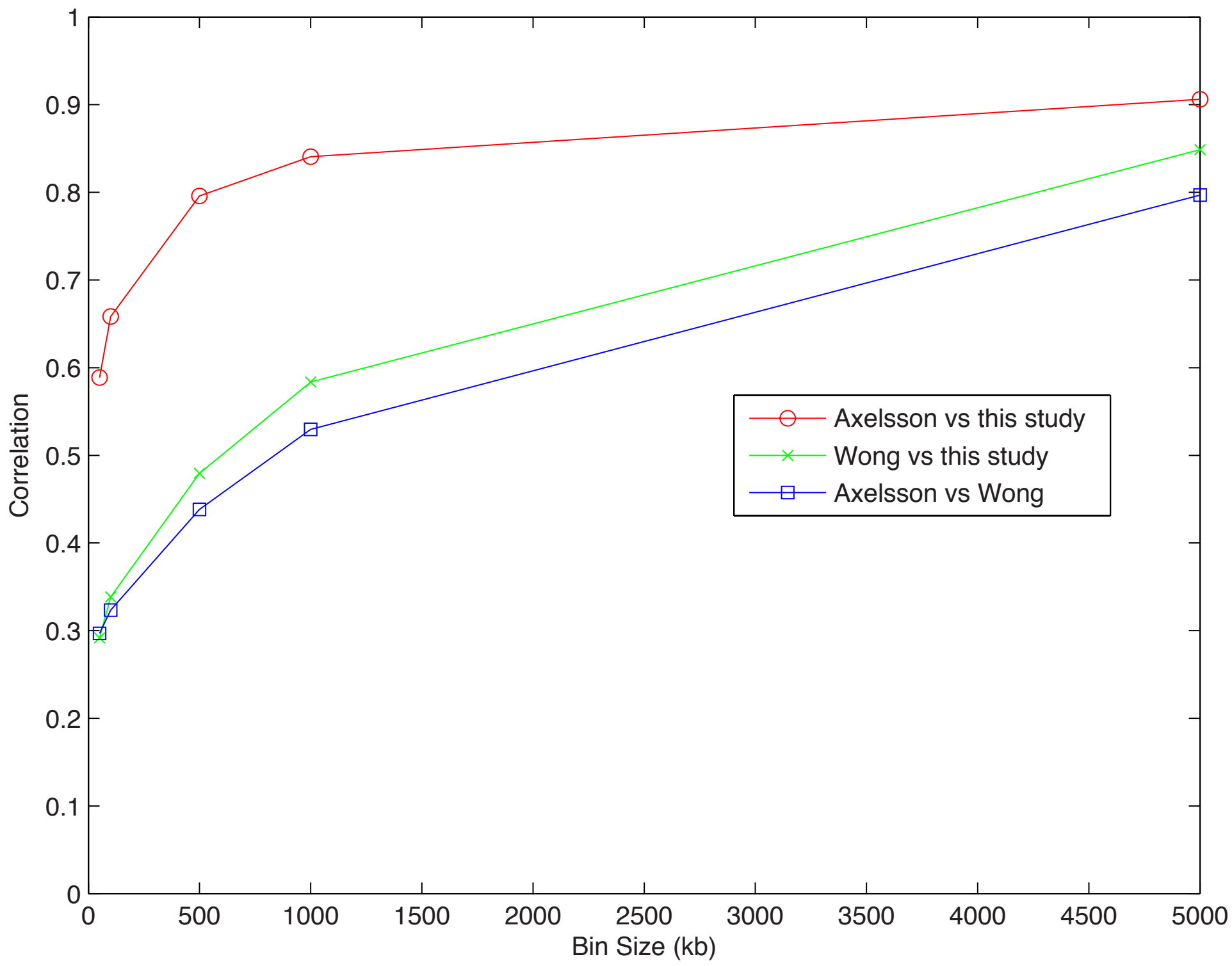

Supplementary Figure 6

A

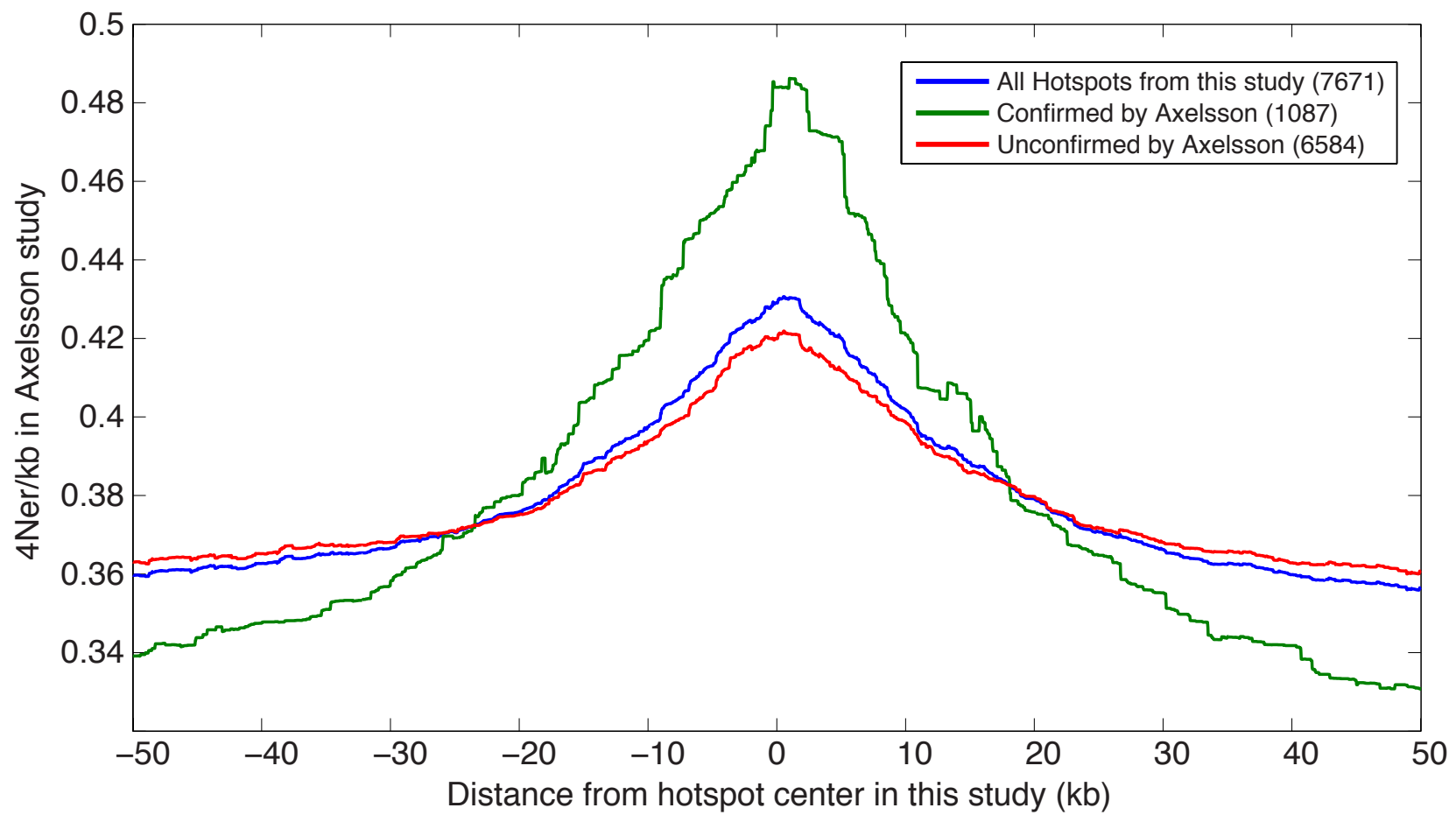

B

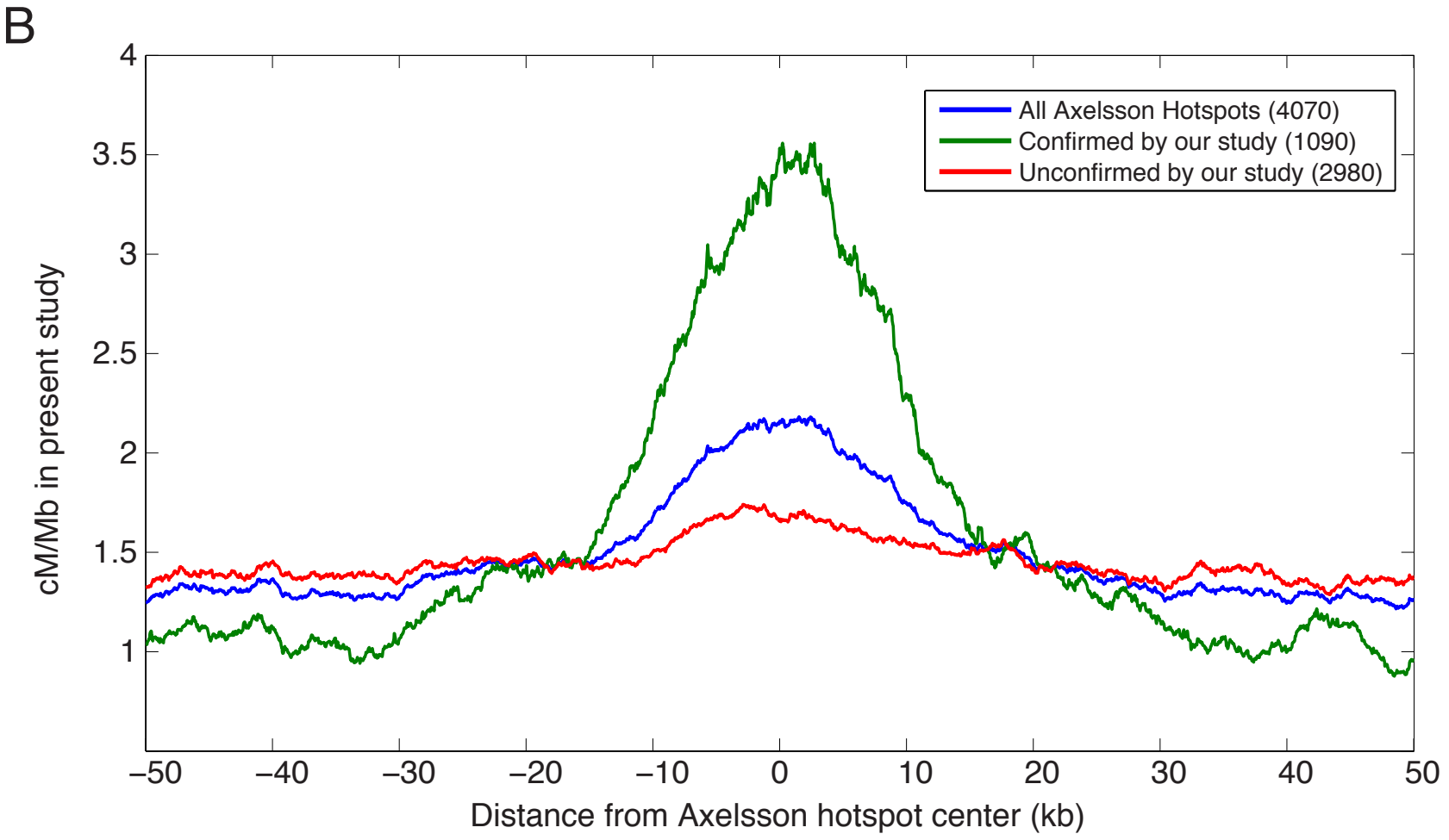

Supplementary Figure 7

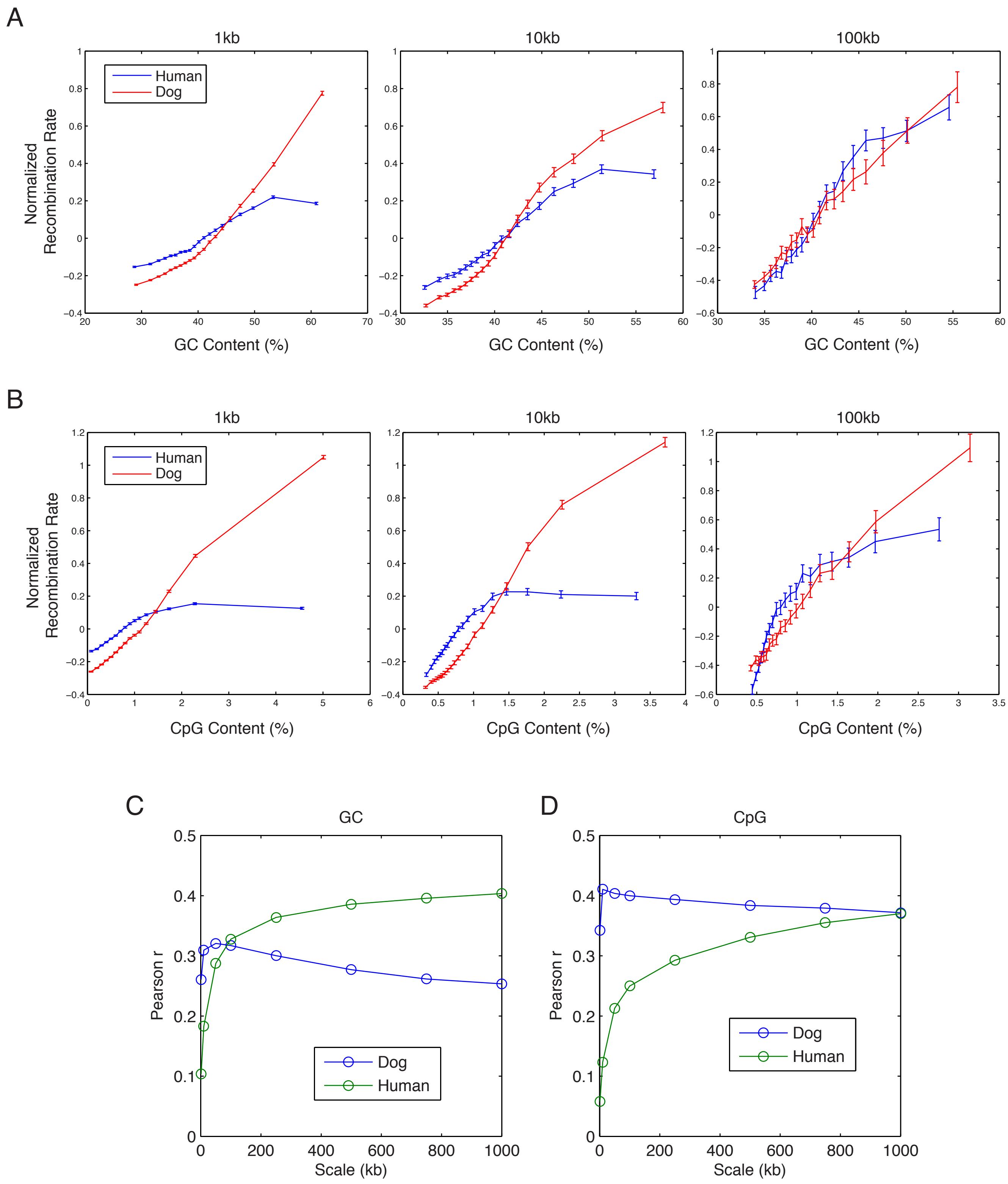

Supplementary Figure 8

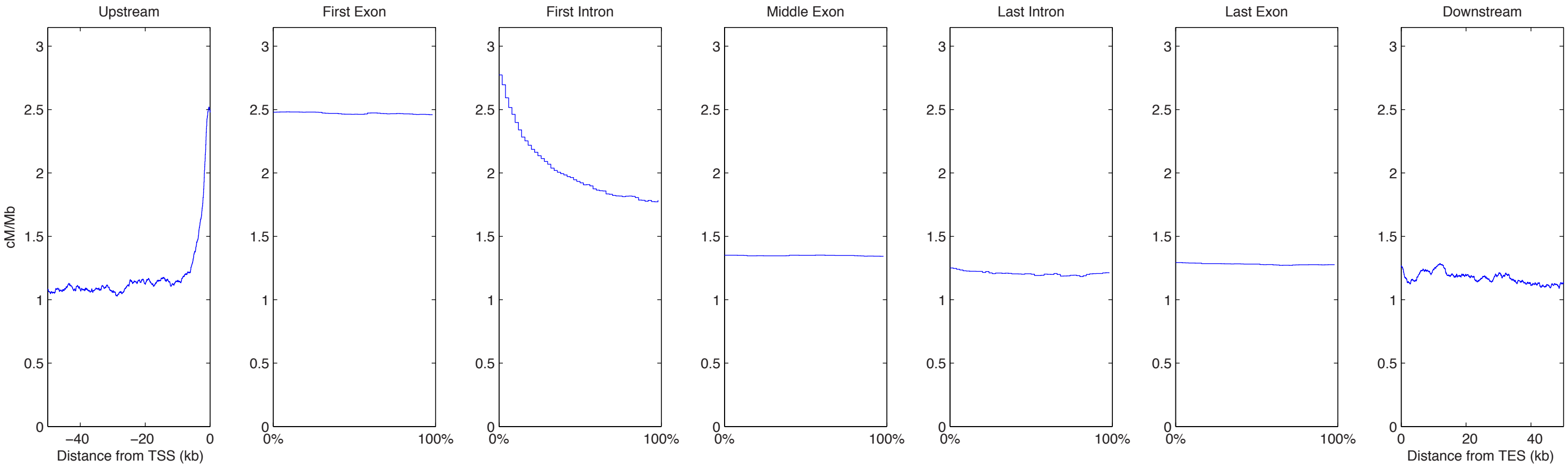

Supplementary Figure 9

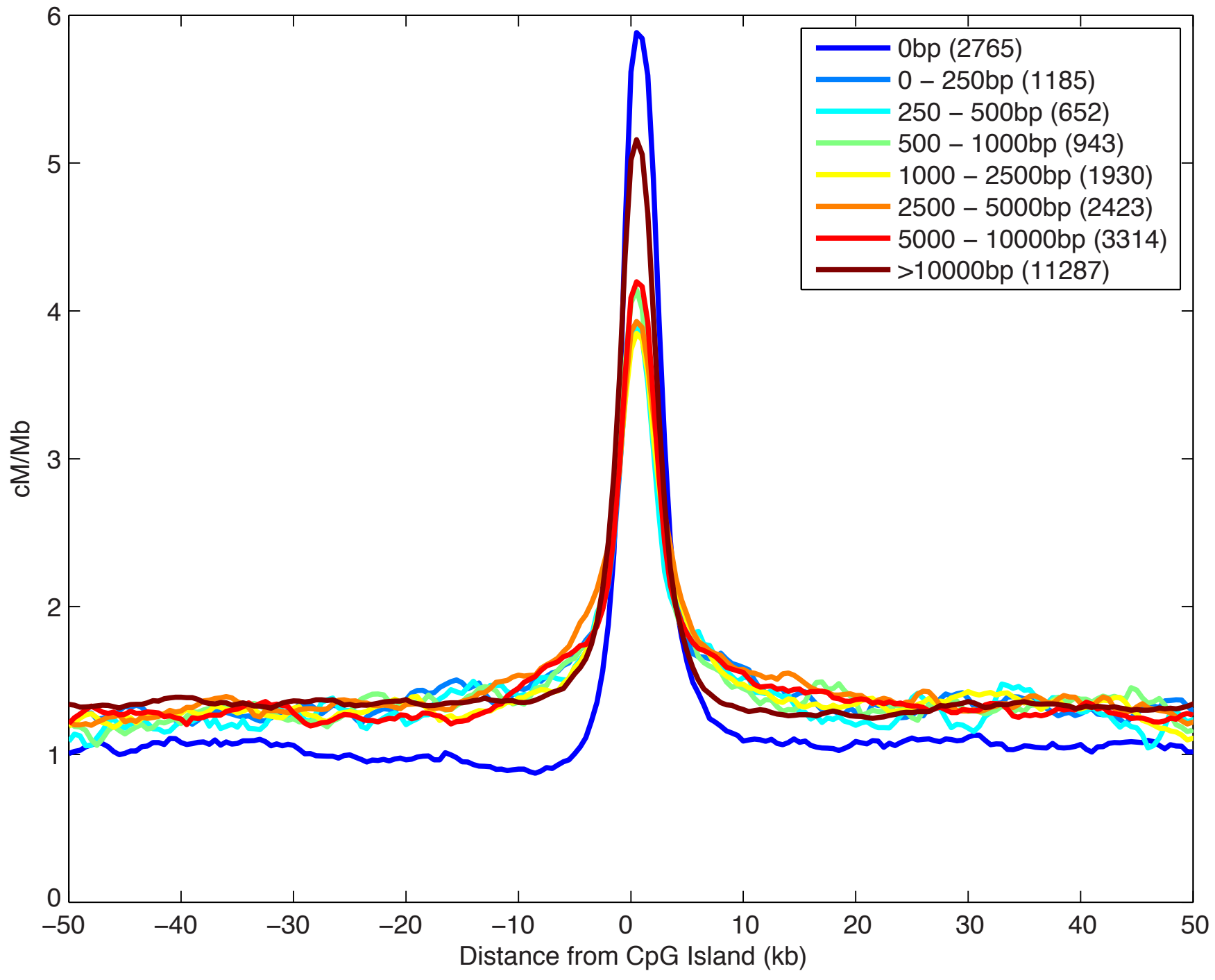

Supplementary Figure 10

A

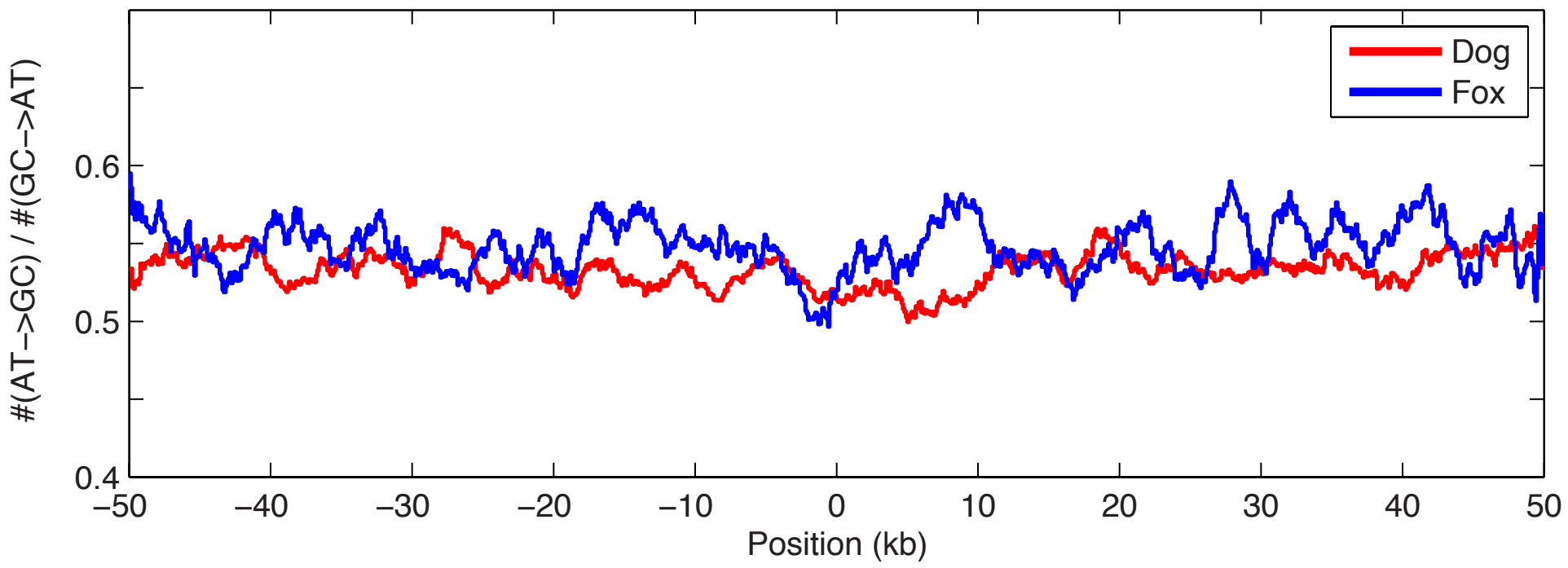

B

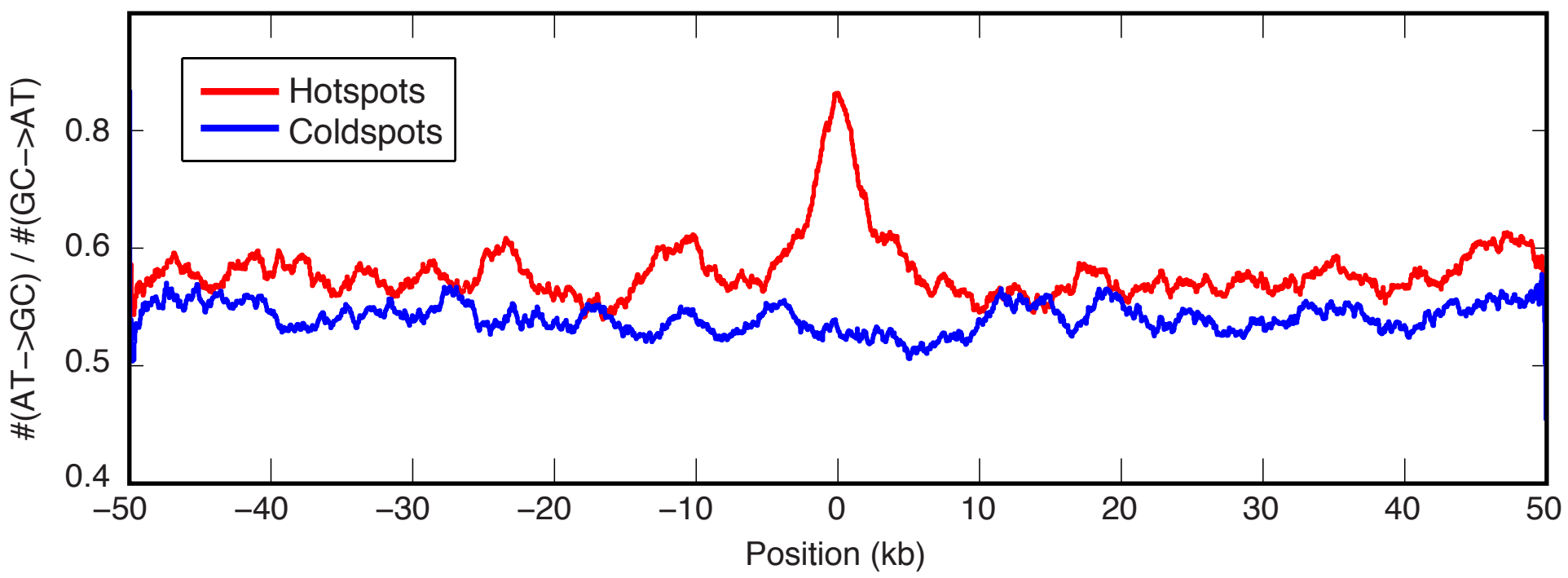

Supplementary Figure 11

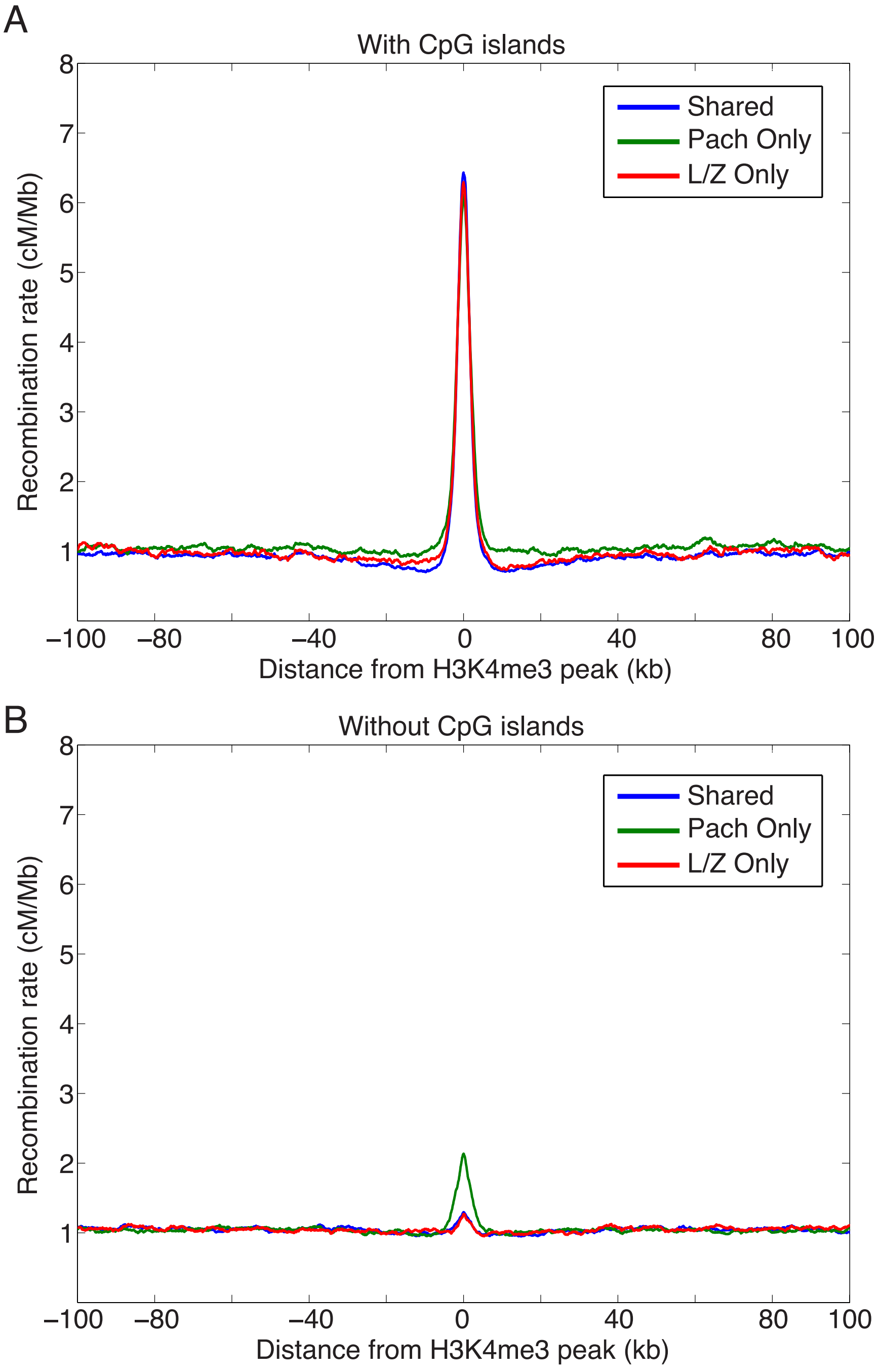

Supplementary Figure 12

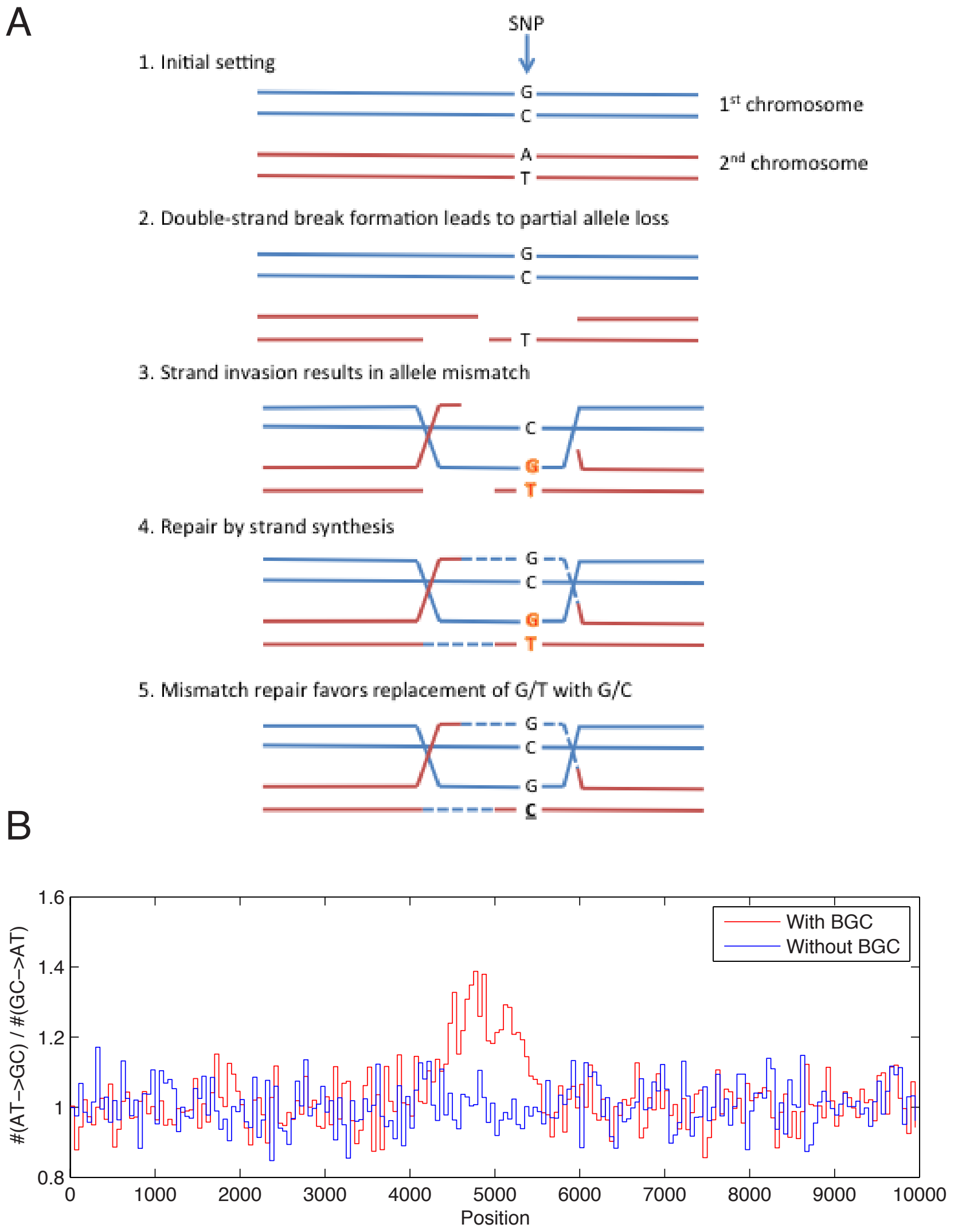

Supplementary Figure 13

**Supplementary Table 1: Locations of collected samples**

| Population/region | N | Individuals |
|---|---|---|
| China (north) | 8 | Xilin Basset, Chow chow, Shar pei, Sichuan Liangshan dog, Sichuan Qingchuan dog, Chongqing dog, Guizhou Xiasi dog, Chinese field dog |
| China (south) | 6 | Kunming dog, Shandong canine, Shaanxi canine, Hebei canine, Mongolia canine, Kazakhstan shepherd dog |
| Europe | 6 | Labrador retriever, Caucasian shepherd, Portugal village dog (2), Croatia village dog, Bosnia village dog |
| India | 6 | Tibetan mastiff/shepherd mix, village dog (5) |
| Mideast | 8 | Afghan hound, Egypt village dog (2), Lebanon village dog (3), Qatar village dog (2) |
| Oceania | 7 | Borneo village dog (3), Papua New Guinea village dog (3), Taiwan village dog |
| Vietnam | 6 | Village dog (6) |
| Other | 4 | Xoloitzcuintl, Namibia village dog (3) |

**Supplementary Table 2: Details of sequence coverage**

| Dog ID | Population | Mean Autosomal Coverage |
|---|---|---|
| 1735 | Afghan Hound | 6.8948 |
| 2972 | Labrador Retriever | 9.0459 |
| 4669 | Xoloitzcuintli | 13.9738 |
| BA19 | Bosnia (Tornjak) | 7.0983 |
| Dog01 | China | 7.4051 |
| Dog02 | China | 8.1309 |
| Dog03 | China | 7.9728 |
| Dog04 | China | 7.4055 |
| Dog06 | China | 7.892 |
| Dog07 | China | 6.6243 |
| Dog08 | China | 7.7976 |
| Dog09 | China | 7.4147 |
| Dog10 | China | 7.8545 |
| Dog11 | China | 7.4648 |
| Dog12 | China | 7.8856 |
| Dog13 | China | 8.2172 |
| Dog14 | China | 7.9076 |
| Dog15 | China | 7.701 |
| EG44 | Egypt | 7.7856 |
| EG49 | Egypt | 7.01 |
| HR85 | Bosnia (Istrian Shorthaired Hound) | 7.7331 |
| HR93 | Bosnia (Caucasian Ovcharka) | 9.1283 |
| ID125 | India | 12.6618 |
| ID137 | India | 7.1976 |
| ID165 | India | 9.469 |
| ID168 | India (Tibetan Masiff mix) | 8.3447 |
| ID60 | India | 13.022 |
| ID91 | India | 7.5009 |
| IN18 | Borneo | 4.9445 |
| IN23 | Borneo | 6.948 |
| IN29 | Borneo | 5.7439 |
| LB74 | Lebanon | 8.3697 |
| LB79 | Lebanon | 8.7643 |
| LB85 | Lebanon | 8.609 |
| NA63 | Namibia | 7.6241 |
| NA8 | Namibia | 12.7724 |
| NA89 | Namibia | 7.0827 |
| PG115 | Papua New Guinea | 8.0253 |
| PG122 | Papua New Guinea | 5.7505 |
| PG84 | Papua New Guinea | 15.0014 |
| PT61 | Portugal | 15.6798 |
| PT71 | Portugal | 13.7652 |
| QA27 | Qatar | 12.382 |
| QA5 | Qatar | 6.0629 |
| TW04 | Taiwan | 14.9396 |
| VN21 | Vietnam | 9.0436 |
| VN37 | Vietnam | 6.7708 |
| VN4 | Vietnam | 9.1031 |
| VN42 | Vietnam | 9.0757 |
| VN59 | Vietnam | 8.6369 |
| VN76 | Vietnam | 6.111 |

**Supplementary Table 3: Predictors of recombination rate**

**Correlation (Pearson coefficient, r)**

| | GC content | CpG content | Genes* |
|---|---|---|---|
| 1kb | 0.2571 | 0.3428 | -0.0284 |
| 10kb | 0.3082 | 0.4104 | -0.0688 |
| 100kb | 0.3162 | 0.3967 | -0.0616 |
| 1Mb | 0.2534 | 0.368 | -0.0679 |

**Multiple Regression with GC and CpG content**

| **Coefficient (b)** | constant | GC | CpG |
|---|---|---|---|
| 1kb | 5.43E-04 | -6.71E-07 | 5.41E-04 |
| 10kb | 0.0139 | -0.0003 | 0.0089 |
| 100kb | 0.1376 | -0.0026 | 0.0731 |
| 1Mb | 2.6939 | -0.0627 | 0.9279 |
| | | | |
| **p-value** | constant | GC | CpG |
| 1kb | 0 | 0.0071 | 0 |
| 10kb | 0 | 1.83E-264 | 0 |
| 100kb | 3.23E-62 | 9.69E-27 | 0.00E+00 |
| 1Mb | 2.10E-27 | 1.12E-17 | 1.03E-54 |

**Multiple Regression with GC, CpG, and gene content**

| **Coefficient (b)** | constant | GC | CpG | Genes* |
|---|---|---|---|---|
| 1kb | 5.53E-04 | -1.55E-07 | 5.41E-04 | -1.19E-04 |
| 10kb | 0.0142 | -0.0003 | 0.009 | -0.0039 |
| 100kb | 0.1367 | -0.0023 | 0.0741 | -0.0388 |
| 1Mb | 2.2791 | -0.047 | 0.8654 | -0.6019 |
| | | | | |
| **p-value** | constant | GC | CpG | Genes* |
| 1kb | 0 | 0.5347 | 0 | 0 |
| 10kb | 0 | 4.33E-236 | 0 | 0 |
| 100kb | 1.71E-62 | 1.86E-22 | 0.00E+00 | 1.00E-84 |
| 1Mb | 2.39E-19 | 6.71E-10 | 9.30E-48 | 1.24E-10 |

**Multiple Regression with GC, GC^2, CpG, and gene content**

| **Coefficient (b)** | constant | GC | GC^2 | CpG | Genes* |
|---|---|---|---|---|---|
| 1kb | -0.0015 | 9.96E-05 | -1.25E-06 | 6.48E-04 | -1.15E-04 |
| 10kb | -0.0522 | 0.0029 | -3.89E-05 | 0.0113 | -0.0038 |
| 100kb | -0.6816 | 0.0369 | -4.77E-04 | 0.1017 | -0.0423 |
| 1Mb | -6.2314 | 0.3676 | -0.0052 | 1.1839 | -0.7575 |
| | | | | | |
| **p-value** | constant | GC | GC^2 | CpG | Genes* |
| 1kb | 0 | 0 | 0 | 0 | 0 |
| 10kb | 0 | 0 | 0 | 0 | 0 |
| 100kb | 5.07E-124 | 5.67E-165 | 1.70E-191 | 0 | 9.07E-104 |
| 1Mb | 2.62E-12 | 5.02E-18 | 4.67E-23 | 9.71E-69 | 3.91E-16 |

* Fraction of bin between transcription start end sites

**Supplementary Table 4: PRDM9 exon 7 DNA sequences**

Positions 1–78

```
JF750638.1|Canis_lupus_familiaris_Beagle                                 CCTAACCCAGAGATACATCCATGTCCATCCTGCTCTCTG-CCTTCTCCAGTCAGAAATTCCTCAGCCAGCATTTGGAA
JF750639.1|Canis_lupus_familiaris_Portugese_Water_Dog                    CCTAACCCAGAGATACATCCATGTCCATCCTGCTCTCTG-CCTTCTCCAGTCAGAAATTCCTCAGCCAGCATTTGGAA
JF750640.1|Canis_lupus_familiaris_Bearded_Collie                         CCTAACCCAGAGATACATCCATGTCCATCCTGCTCTCTG-CCTTCTCCAGTCAGAAATTCCTCAGCCAGCATTTGGAA
JF750641.1|Canis_lupus_familiaris_Standard_Poodle                        CCTAACCCAGAGATACATCCATGTCCATCCTGCTCTCTG-CCTTCTCCAGTCAGAAATTCCTCAGCCAGCATTTGGAA
JF750642.1|Canis_lupus_familiaris_Boxer                                  CCTAACCCAGAGATACATCCATGTCCATCCTGCTCTCTG-CCTTCTCCAGTCAGAAATTCCTCAGCCAGCATTTGGAA
JF750643.1|Canis_lupus_familiaris_Dalmatian                              CCTAACCCAGAGATACATCCATGTCCATCCTGCTCTCTG-CCTTCTCCAGTCAGAAATTCCTCAGCCAGCATTTGGAA
JF750644.1|Canis_lupus_familiaris_Labrador_retriever                     CCTAACCCAGAGATACATCCATGTCCATCCTGCTCTCTG-CCTTCTCCAGTCAGAAATTCCTCAGCCAGCATTTGGAA
JF750645.1|Canis_lupus_familiaris_German_Shepard                         CCTAACCCAGAGATACATCCATGTCCATCCTGCTCTCTG-CCTTCTCCAGTCAGAAATTCCTCAGCCAGCATTTGGAA
JF750646.1|Canis_lupus_familiaris_Rough_Collie                           CCTAACCCAGAGATACATCCATGTCCATCCTGCTCTCTG-CCTTCTCCAGTCAGAAATTCCTCAGCCAGCATTTGGAA
JF750647.1|Canis_lupus_familiaris_Eurasier                               CCTAACCCAGAGATACATCCATGTCCATCCTGCTCTCTG-CCTTCTCCAGTCAGAAATTCCTCAGCCAGCATTTGGAA
JF750648.1|Canis_lupus_familiaris_Polish_Lowland_Sheepdog                CCTAACCCAGAGATAC????????CCATCCTGCTCTCTG-CCTTCTCCAGTCAGAAATTCCTCAGCCAGCATTTGGAA
JF750649.1|Canis_lupus_familiaris_Rottweiler                             CCTAACCCAGAGATACATCCATGTCCATCCTGCTCTCTG-CCTTCTCCAGTCAGAAATTCCTCAGCC?GCATTTGGAA
JF750650.1|Canis_lupus_familiaris_Smalandsstrovare                       CCTAACCCAGAGATACATCCATGTCCATCCTGCTCTCTG-CCTTCTCCAGTCAGAAATTCCTCAGCCAGCATTTGGAA
JF750651.1|Canis_lupus_familiaris_Swedish_Elkhound                       CCTAACCCAGAGATAC????????CCATCCTGCTCTCTG-CCTTCTCCAGTCAGAAATTCCTCAGCCAGCATTTGGAA
JF750652.1|Canis_lupus_familiaris_Nova_Scotia_Duck_Tolling_Retriever     CCTAACCCAGAGATACATCCATGTCCATCCTGCTCTCTG-CCTTCTCCAGTCAGAAATTCCTCAGCCAGCATTTGGAA
JF750653.1|Canis_lupus_isolate_ref21                                     CCTAACCCAGAGATACATCCATGTCCATCCTGCTCTCTG-CCTTCTCCAGTCAGAAATTCCTCAGCCAGCATTTGGAA
JF750654.1|Canis_lupus_isolate_ref34                                     CCTAACCCAGAGATACATCCATGTCCATCCTGCTCTCTG-CCTTCTCCAGTCAGAAATTCCTCAGCCAGCATTTGGAA
JF750655.1|Vulpes_vulpes                                                 CCTAACCCAGAGATATATCCATGTCCATCCTGCTCTCTGTCCTTCTCCAGTCAGAAATTCCTCAGCCAGCATTTGGAA
JF750656.1|Speothos_venaticus                                            CCTAACCCAGAGATACATCCATGTCCATCCTGCTCTCTG-CCTTCTCCAGTCAGAAATTCCTCAGCCAGCATTTGGAA
JF750657.1|Lycaon_pictus                                                 CCTAACCCAGAGATACATCCATGTCCATCCTGCTCTCTG-CCTTCTCCAGTCAGAAATTCCTCAGCCAGCATTTGGAA
JF750658.1|Canis_mesomelas                                               CCTAACCCAGAGATACATCCATGTCCATCCTGCTCTCTG-CCTTCTCCAGTCAGAAATTCCTCAGCCAGCATTTGGAA
JF750659.1|Canis_aureus                                                  CCTAACCCAGAGATACATCCATGTCCATCCTGCTCTCTG-CCTTCTCCAGTCAGAAATTCCTCAGCCAGCATTTGGAA
Island_Fox|Urocyon littoralis                                            CYTAACCCAGAGATACATCCATGTCCATCCTGCTCTCTGGCCTTCTCCAGTCAGAAATTCCTCAGCCAGCATTTGGAA
Culpeo|Lycalopex culpaeus                                                CCTAACCCAGAGATACATCCATGTCCATCCTGCTCTCTG-CCTTCTCCAGTCAGAAATTCCTCAGCCAGCATTTGGAA
```

Positions 79–156

```
JF750638.1|Canis_lupus_familiaris_Beagle                                 CACAAYCACCCTTCTCAGATCCTCCCACGAATATCTGTAAGAGAACATTTCCGACCAAAAGATCCATGCCCAGGTTGT
JF750639.1|Canis_lupus_familiaris_Portugese_Water_Dog                    CACAATCACCCTTCTCAGATCCTCCCACGAATATCTGTAAGAGAACATTTCCGACCAAAAGATCCATGCCCAGGTTGT
JF750640.1|Canis_lupus_familiaris_Bearded_Collie                         CACAATCACCCTTCTCAGATCCTCCCACGAATATCTGTAAGAGAACATTTCCGACCAAAAGATCCATGCCCAGGTTGT
JF750641.1|Canis_lupus_familiaris_Standard_Poodle                        CACAATCACCCTTCTCAGATCCTCCCACGAATATCTGTAAGAGAACATTTCCGACCAAAAGATCCATGCCCAGGTTGT
JF750642.1|Canis_lupus_familiaris_Boxer                                  CACAATCACCCTTCTCAGATCCTCCCACGAATATCTGTAAGAGAACATTTCCGACCAAAAGATCCATGCCCAGGTTGT
JF750643.1|Canis_lupus_familiaris_Dalmatian                              CACAATCACCCTTCTCAGATCCTCCCACGAATATCTGTAAGAGAACATTTCCGACCAAAAGATCCATGCCCAGGTTGT
JF750644.1|Canis_lupus_familiaris_Labrador_retriever                     CACAATCACCCTTCTCAGATCCTCCCACGAATATCTGTAAGAGAACATTTCCGACCAAAAGATCCATGCCCAGGTTGT
JF750645.1|Canis_lupus_familiaris_German_Shepard                         CACAACCACCCTTCTCAGATCCTCCCACGAATATCTGTAAGAGAACATTTCCGACCAAAAGATCCATGCCCAGGTTGT
JF750646.1|Canis_lupus_familiaris_Rough_Collie                           CACAATCACCCTTCTCAGATCCTCCCACGAATATCTGTAAGAGAACATTTCCGACCAAAAGATCCATGCCCAGGTTGT
JF750647.1|Canis_lupus_familiaris_Eurasier                               CACAATCACCCTTCTCAGATCCTCCCACGAATATCTGTAAGAGAACATTTCCGACCAAAAGATCCATGCCCAGGTTGT
JF750648.1|Canis_lupus_familiaris_Polish_Lowland_Sheepdog                CACAATCACCCTTCTCAGATCCTCCCACGAATATCTGTAAGAGAACATTTCCGACCAAAAGATCCATGCCCAGGTTGT
JF750649.1|Canis_lupus_familiaris_Rottweiler                             CACAAYCACCCTTCTCAGATCCTCCCACGAATATCTGTAAGAGAACATTTCCGACCAAAAGATCCATGCCCAGGTTGT
JF750650.1|Canis_lupus_familiaris_Smalandsstrovare                       CACAATCACCCTTCTCAGATCCTCCCACGAATATCTGTAAGAGAACATTTCCGACCAAAAGATCCATGCCCAGGTTGT
JF750651.1|Canis_lupus_familiaris_Swedish_Elkhound                       CACAAYCACCCTTCTCAGATCCTCCCACGAATATCTGTAAGAGAACATTTCCGACCAAAAGATCCATGCCCAGGTTGT
JF750652.1|Canis_lupus_familiaris_Nova_Scotia_Duck_Tolling_Retriever     CACAATCACCCTTCTCAGATCCTCCCACGAATATCTGTAAGAGAACATTTCCGACCAAAAGATCCATGCCCAGGTTGT
JF750653.1|Canis_lupus_isolate_ref21                                     CACAATCACCCTTCTCAGATCCTCCCACGAATATCTGTAAGAGAACATTTCCGACCAAAAGATCCATGCCCAGGTTGT
JF750654.1|Canis_lupus_isolate_ref34                                     CACAATCACCCTTCTCAGATCCTCCCACGAATATCTGTAAGAGAACATTTCCGACCAAAAGATCCATGCCCAGGTTGT
JF750655.1|Vulpes_vulpes                                                 CACAATCACCCTTCTCAGATCCTCCCACGAATATCTATAAGAGAACATTTCCAACCAAAAGATCCATGCCCAGGTTGT
JF750656.1|Speothos_venaticus                                            CACAATCACCCTTCTCAGATCCTCCCATGAATATCTGTAAGAGAACATTTCCGACCAAAAGATCCATGCCCAGGTTGT
JF750657.1|Lycaon_pictus                                                 CACAATCACCCTTCTCAGATCCTCCCACGAATATCTGTAAGAGAACATTTCCGACCAAAAGATCCATGCCCAGGTTGT
JF750658.1|Canis_mesomelas                                               CACAATCACCCTTCTCAGATCCTCCCACAAATATCTGTAAGAGAACATTTCCGACCAAAAGATCCATGCCCAGGTTGT
JF750659.1|Canis_aureus                                                  CACAATCACCCTTCTCAGATCCTCCCACGAATATCTGTAAGAGAACATTTCCGACCAAAAGATCCATGCCCAGGTTGT
Island_Fox|Urocyon littoralis                                            CACAATCACCCTTMTCAGATCCTCCCACGAATATCTGTAAGAGAACATTTCCGACCAAAAGATCCATGCCCAGGTTGT
Culpeo|Lycalopex culpaeus                                                CACAATCACCCTTCTCAGATCCTCCCACGAATATCTGTAAGAGAACATTTCTGACCAAAAGATCCATGCCCAGGTTGT
```

Positions 157–234

```
JF750638.1|Canis_lupus_familiaris_Beagle                                 CAGAATCAGCAGCAGCAGCAACATTCTGATCCACAGCGCTGGAATGACAGAGCTAAAGGTCAAGAGGGCAAAGAAAGG
JF750639.1|Canis_lupus_familiaris_Portugese_Water_Dog                    CAGAATCAGCAGCAGCAGCAACATTCTGATCCACAGCGCTGGAATGACAGAGCTAAAGGTCAAGAGGGCAAAGAAAGG
JF750640.1|Canis_lupus_familiaris_Bearded_Collie                         CAGAATCAGCAGCAGCAGCAACATTCTGATCCACAGCGCTGGAATGACAGAGCTAAAGGTCAAGAGGGCAAAGAAAGG
JF750641.1|Canis_lupus_familiaris_Standard_Poodle                        CAGAATCAGCAGCAGCAGCAACATTCTGATCCACAGCGCTGGAATGACAGAGCTAAAGGTCAAGAGGGCAAAGAAAGG
JF750642.1|Canis_lupus_familiaris_Boxer                                  CAGAATCAGCAGCAGCAGCAACATTCTGATCCACAGCGCTGGAATGACAGAGCTAAAGGTCAAGAGGGCAAAGAAAGG
JF750643.1|Canis_lupus_familiaris_Dalmatian                              CAGAATCAGCAGCAGCAGCAACATTCTGATCCACAGCGCTGGAATGACAGAGCTAAAGGTCAAGAGGGCAAAGAAAGG
JF750644.1|Canis_lupus_familiaris_Labrador_retriever                     CAGAATCAGCAGCAGCAGCAACATTCTGATCCACAGCGCTGGAATGACAGAGCTAAAGGTCAAGAGGGCAAAGAAAGG
JF750645.1|Canis_lupus_familiaris_German_Shepard                         CAGAATCAGCAGCAGCAGCAACATTCTGATCCACAGCGCTGGAATGACAGAGCTAAAGGTCAAGAGGGCAAAGAAAGG
JF750646.1|Canis_lupus_familiaris_Rough_Collie                           CAGAATCAGCAGCAGCAGCAACATTCTGATCCACAGCGCTGGAATGACAGAGCTAAAGGTCAAGAGGGCAAAGAAAGG
JF750647.1|Canis_lupus_familiaris_Eurasier                               CAGAATCAGCAGCAGCAGCAACATTCTGATCCACAGCGCTGGAATGACAGAGCTAAAGGTCAAGAGGGCAAAGAAAGG
JF750648.1|Canis_lupus_familiaris_Polish_Lowland_Sheepdog                CAGAATCAGCAGCAGCAGCAACATTCTGATCCACAGCGCTGGAATGACAGAGCTAAAGGTCAAGAGGGCAAAGAAAGG
JF750649.1|Canis_lupus_familiaris_Rottweiler                             CAGAATCAGCAGCAGCAGCAACATTCTGATCCACAGCGCTGGAATGACAGAGCTAAAGGTCAAGAGGGCAAAGAAAGG
JF750650.1|Canis_lupus_familiaris_Smalandsstrovare                       CAGAATCAGCAGCAGCAGCAACATTCTGATCCACAGCGCTGGAATGACAGAGCTAAAGGTCAAGAGGGCAAAGAAAGG
JF750651.1|Canis_lupus_familiaris_Swedish_Elkhound                       CAGAATCAGCAGCAGCAGCAACATTCTGATCCACAGCGCTGGAATGACAGAGCTAAAGGTCAAGAGGGCAAAGAAAGG
JF750652.1|Canis_lupus_familiaris_Nova_Scotia_Duck_Tolling_Retriever     CAGAATCAGCAGCAGCAGCAACATTCTGATCCACAGCGCTGGAATGACAGAGCTAAAGGTCAAGAGGGCAAAGAAAGG
JF750653.1|Canis_lupus_isolate_ref21                                     CAGAATCAGCAGCAGCAGCAACATTCTGATCCACAGCGCTGGAATGACAGAGCTAAAGGTCAAGAGGGCAAAGAAAGG
JF750654.1|Canis_lupus_isolate_ref34                                     CAGAATCAGCAGCAGCAGCAACATTCTGATCCACAGCGCTGGAATGACAGAGCTAAAGGTCAAGAGGGCAAAGAAAGG
JF750655.1|Vulpes_vulpes                                                 CAGAATCAGCAGCAGCAGCAACATTCTGATCCACAGTGCTGGAATGACAGAGCTAAAGGTCAAGAGGGCAAAGAAAGG
JF750656.1|Speothos_venaticus                                            CAGAATCAGCAGCAGCAGCAACATTCTGATCCACAGCGCTGGAATGACAGAGCTAAAGGTCAAGAGGGCAAAGAAAGG
JF750657.1|Lycaon_pictus                                                 CAGAATCAGCAGCAGCAGCAACATTCTGATCCACAGCGCTGGAGTGACAGAGCTAAAGGTCAAGAGGGCAAAGAAAGG
JF750658.1|Canis_mesomelas                                               CAGAATCAGCAGCAGCAGCAACATTCTGATCCACAGCGCTGGAATGACAGAGCTAAAGGTCAAGAGGGCAAAGAAAGG
JF750659.1|Canis_aureus                                                  CAGAATCAGCAGCAGCAGCAACATTCTGATCCACAGCGCTGGAATGACAGAGCTAAAGGTCAAGAGGGCAAAGAAAGG
Island_Fox|Urocyon littoralis                                            CAGAATCAGCAGCAGCAGCAACATTCTGATCCACAGCGCTGGAATGACAGAGCTAAAGGTCAAGAGGGCAAAGAAAGG
Culpeo|Lycalopex culpaeus                                                CAGAATCAGCAGCAGCAGCAACATTCTGATCCACAGCGCTGGAATGACAGAGCTAAAGGTCAAGAGGGCAAAGAAAGG
```

Positions 235–312

```
JF750638.1|Canis_lupus_familiaris_Beagle                                 TTCAAACCTTTGCCTAAAAGTATAAGGCAGAGAAGAATTTCAAGGGCCTTTTCCACCCCGTGCAAAGGGCAAACAACG
JF750639.1|Canis_lupus_familiaris_Portugese_Water_Dog                    TTCAAACCTTTGCCTAAAAGTATAAGGCAGAGAAGAATTTCAAGGGCCTTTTCCACCCCGTGCAAAGGGCAAACAACG
JF750640.1|Canis_lupus_familiaris_Bearded_Collie                         TTCAAACCTTTGCCTAAAAGTATAAGGCAGAGAAGAATTTCAAGGGCCTTTTCCACCCCGTGCAAAGGGCAAACAACG
JF750641.1|Canis_lupus_familiaris_Standard_Poodle                        TTCAAACCTTTGCCTAAAAGTATAAGGCAGAGAAGAATTTCAAGGGCCTTTTCCACCCCGTGCAAAGGGCAAACAACG
JF750642.1|Canis_lupus_familiaris_Boxer                                  TTCAAACCTTTGCCTAAAAGTATAAGGCAGAGAAGAATTTCAAGGGCCTTTTCCACCCCGTGCAAAGGGCAAACAACG
JF750643.1|Canis_lupus_familiaris_Dalmatian                              TTCAAACCTTTGCCTAAAAGTATAAGGCAGAGAAGAATTTCAAGGGCCTTTTCCACCCCGTGCAAAGGGCAAACAACG
JF750644.1|Canis_lupus_familiaris_Labrador_retriever                     TTCAAACCTTTGCCTAAAAGTATAAGGCAGAGAAGAATTTCAAGGGCCTTTTCCACCCCGTGCAAAGGGCAAACAACG
JF750645.1|Canis_lupus_familiaris_German_Shepard                         TTCAAACCTTTGCCTAAAAGTATAAGGCAGAGAAGAATTTCAAGGGCCTTTTCCACCCCGTGCAAAGGGCAAACAACG
JF750646.1|Canis_lupus_familiaris_Rough_Collie                           TTCAAACCTTTGCCTAAAAGTATAAGGCAGAGAAGAATTTCAAGGGCCTTTTCCACCCCGTGCAAAGGGCAAACAACG
JF750647.1|Canis_lupus_familiaris_Eurasier                               TTCAAACCTTTGCCTAAAAGTATAAGGCAGAGAAGAATTTCAAGGGCCTTTTCCACCCCGTGCAAAGGGCAAACAACG
JF750648.1|Canis_lupus_familiaris_Polish_Lowland_Sheepdog                TTCAAACCTTTGCCTAAAAGTATAAGGCAGAGAAGAATTTCAAGGGCCTTTTCCACCCCGTGCAAAGGGCAAACAACG
JF750649.1|Canis_lupus_familiaris_Rottweiler                             TTCAAACCTTTGCCTAAAAGTATAAGGCAGAGAAGAATTTCAAGGGCCTTTTCCACCCCGTGCAAAGGGCAAACAACG
JF750650.1|Canis_lupus_familiaris_Smalandsstrovare                       TTCAAACCTTTGCCTAAAAGTATAAGGCAGAGAAGAATTTCAAGGGCCTTTTCCACCCCGTGCAAAGGGCAAACAACG
JF750651.1|Canis_lupus_familiaris_Swedish_Elkhound                       TTCAAACCTTTGCCTAAAAGTATAAGGCAGAGAAGAATTTCAAGGGCCTTTTCCACCCCGTGCAAAGGGCAAACAACG
JF750652.1|Canis_lupus_familiaris_Nova_Scotia_Duck_Tolling_Retriever     TTCAAACCTTTGCCTAAAAGTATAAGGCAGAGAAGAATTTCAAGGGCCTTTTCCACCCCGTGCAAAGGGCAAACAACG
JF750653.1|Canis_lupus_isolate_ref21                                     TTCAAACCTTTGCCTAAAAGTATAAGGCAGAGAAGAATTTCAAGGGCCTTTTCCACCCCGTGCAAAGGGCAAACAACG
JF750654.1|Canis_lupus_isolate_ref34                                     TTCAAACCTTTGCCTAAAAGTATAAGGCAGAGAAGAATTTCAAGGGCCTTTTCCACCCCGTGCAAAGGGCAAACAACG
JF750655.1|Vulpes_vulpes                                                 TTCAAACCTTTGCCTAAA-GTATAAGGCAGAGAAAAATTTCAAGGGCCTTTTCCACCCCGTGCAAAGGACAAACAACG
JF750656.1|Speothos_venaticus                                            TTCAAACCTTTGCCTAAACGTATAAGGCAGAGAAGAATTTCAAGGGCCTTTTCCACCCCGTGCAAAGGGCAAACAACG
JF750657.1|Lycaon_pictus                                                 TTCAAACCTTTGCCTAAAAGTATAAGGCAGAGAAGAATTTCAAGGGCCTTTTCCACCCCGTGCAAAGGGCAAACAACG
JF750658.1|Canis_mesomelas                                               TTCAAACCTTTGCCTAAAAGTATAAGGCAGAGAAGAATTTCAAGGGCCTTTTCCACCCCGTGCAAAGGGCAAACAACG
JF750659.1|Canis_aureus                                                  TTCAAACCTTTGCCTAAAAGTATAAGGCAGAGAAGAATTTCAAGGGCCTTTTCCACCCCGTGCAAAGGGCAAACAACG
Island_Fox|Urocyon littoralis                                            TTCAAACCTTTGCCTAAAAGTATAAGGCAGAGAAGAATTTCAAGGGCCTTTTCTACCCCGTGCAAAGGACAAACAACG
Culpeo|Lycalopex culpaeus                                                TTCAAACCTTTGCCTAAACGTATAAGGCAGAGAAGAATTTCAAGAGCCTTTTCCACCCCGTGCAAAGGGCAAACAACG
```

Positions 313–390

```
JF750638.1|Canis_lupus_familiaris_Beagle                                 TGTGAGGGAATAGTGAAGGAAGAGCCCAGCGCAGGCTCACAGAAATTGAATCCAGAGGACACAGGCAAATTATTCAAG
JF750639.1|Canis_lupus_familiaris_Portugese_Water_Dog                    TGTGAGGGAATAGTGAAGGAAGAGCCCAGCGCAGGCTCACAGAAATTGAATCCAGAGGACACAGGCAAATTATTCAAG
JF750640.1|Canis_lupus_familiaris_Bearded_Collie                         TGTGAGGGAATAGTGAAGGAAGAGCCCAGCGCAGGCTCACAGAAATTGAATCCAGAGGACACAGGCAAATTATTCAAG
JF750641.1|Canis_lupus_familiaris_Standard_Poodle                        TGTGAGGGAATAGTGAAGGAAGAGCCCAGCGCAGGCTCACAGAAATTGAATCCAGAGGACACAGGCAAATTATTCAAG
JF750642.1|Canis_lupus_familiaris_Boxer                                  TGTGAGGGAATAGTGAAGGAAGAGCCCAGCGCAGGCTCACAGAAATTGAATCCAGAGGACACAGGCAAATTATTCAAG
JF750643.1|Canis_lupus_familiaris_Dalmatian                              TGTGAGGGAATAGTGAAGGAAGAGCCCAGCGCAGGCTCACAGAAATTGAATCCAGAGGACACAGGCAAATTATTCAAG
JF750644.1|Canis_lupus_familiaris_Labrador_retriever                     TGTGAGGGAATAGTGAAGGAAGAGCCCAGCGCAGGCTCACAGAAATTGAATCCAGAGGACACAGGCAAATTATTCAAG
JF750645.1|Canis_lupus_familiaris_German_Shepard                         TGTGAGGGAATAGTGAAGGAAGAGCCCAGCGCAGGCTCACAGAAATTGAATCCAGAGGACACAGGCAAATTATTCAAG
JF750646.1|Canis_lupus_familiaris_Rough_Collie                           TGTGAGGGAATAGTGAAGGAAGAGCCCAGCGCAGGCTCACAGAAATTGAATCCAGAGGACACAGGCAAATTATTCAAG
JF750647.1|Canis_lupus_familiaris_Eurasier                               TGTGAGGGAATAGTGAAGGAAGAGCCCAGCGCAGGCTCACAGAAATTGAATCCAGAGGACACAGGCAAATTATTCAAG
JF750648.1|Canis_lupus_familiaris_Polish_Lowland_Sheepdog                TGTGAGGGAATAGTGAAGGAAGAGCCCAGCGCAGGCTCACAGAAATTGAATCCAGAGGACACAGGCAAATTATTCAAG
JF750649.1|Canis_lupus_familiaris_Rottweiler                             TGTGAGGGAATAGTGAAGGAAGAGCCCAGCGCAGGCTCACAGAAATTGAATCCAGAGGACACAGGCAAATTATTCAAG
JF750650.1|Canis_lupus_familiaris_Smalandsstrovare                       TGTGAGGGAATAGTGAAGGAAGAGCCCAGCGCAGGCTCACAGAAATTGAATCCAGAGGACACAGGCAAATTATTCAAG
JF750651.1|Canis_lupus_familiaris_Swedish_Elkhound                       TGTGAGGGAATAGTGAAGGAAGAGCCCAGCGCAGGCTCACAGAAATTGAATCCAGAGGACACAGGCAAATTATTCAAG
JF750652.1|Canis_lupus_familiaris_Nova_Scotia_Duck_Tolling_Retriever     TGTGAGGGAATAGTGAAGGAAGAGCCCAGCGCAGGCTCACAGAAATTGAATCCAGAGGACACAGGCAAATTATTCAAG
JF750653.1|Canis_lupus_isolate_ref21                                     TGTGAGGGAATAGTGAAGGAAGAGCCCAGCGCAGGCTCACAGAAATTGAATCCAGAGGACACAGGCAAATTATTCAAG
JF750654.1|Canis_lupus_isolate_ref34                                     TGTGAGGGAATAGTGAAGGAAGAGCCCAGCGCAGGCTCACAGAAATTGAATCCAGAGGACACAGGCAAATTATTCAAG
JF750655.1|Vulpes_vulpes                                                 TGTGAGGGAATAGTGAAGGAAGAGCCCAGCGCAGGCTCACAGAAATTGAATCCAGAGGACACAGGCAAATTATTCAAG
JF750656.1|Speothos_venaticus                                            TGTGAGGGAATAGTGAAGGAAGAGCCCAGCGCAAGCTCACAGAAATTGAATCCAGAGGAYACAGGCAAATTATTCAAG
JF750657.1|Lycaon_pictus                                                 TGTGAGGGAATAGTGAAGGAAGAGCCCAGCGCAGGCTCACAGAAATTGAATCCAGAGGACACAGGCAAATTATTCAAG
JF750658.1|Canis_mesomelas                                               TGTGAGGGAATAGTGAAGGAAGAGCCCAGCGCAGGCTCACAGAAATTGAATCCAGAGGACACAGGCAAATTATTCAAG
JF750659.1|Canis_aureus                                                  TGTGAGGGAATAGTGAAGGAAGAGCCCAGCGCAGGCTCACAGAAATTGAATCCAGAGGACACAGGCAAATTATTCAAG
Island_Fox|Urocyon littoralis                                            TGTGAGGGAATAGTGAAGGAAGAGCCCAGCACAGGCTCACAGAAATTGAATCCAGAGGACACAGGCAAATTATTCAAG
Culpeo|Lycalopex culpaeus                                                TGTGAGGGAATAGTGAAGGAAGAGCCCAGCGCAGGCTCACAGAAATTGAATCCAGAGGACACAGGCAAATTATTCAAG
```

Positions 391–468

```
JF750638.1|Canis_lupus_familiaris_Beagle                                 GGGGTAGGAATGACAAGAATTATAAGAGTCAAATATAGAGGCTGTGGGCGAGGCTTCAATGATAGGTCACATCTCAGC
JF750639.1|Canis_lupus_familiaris_Portugese_Water_Dog                    GGGGTAGGAATGACAAGAATTATAAGAGTCAAATATAGAGGCTGTGGGCGAGGCTTCAATGATAGGTCACATCTCAGC
JF750640.1|Canis_lupus_familiaris_Bearded_Collie                         GGGGTAGGAATGACAAGAATTATAAGAGTCAAATATAGAGGCTGTGGGCGAGGCTTCAATGATAGGTCACATCTCAGC
JF750641.1|Canis_lupus_familiaris_Standard_Poodle                        GGGGTAGGAATGACAAGAATTATAAGAGTCAAATATAGAGGCTGTGGGCGAGGCTTCAATGATAGGTCACATCTCAGC
JF750642.1|Canis_lupus_familiaris_Boxer                                  GGGGTAGGAATGACAAGAATTATAAGAGTCAAATATAGAGGCTGTGGGCGAGGCTTCAATGATAGGTCACATCTCAGC
JF750643.1|Canis_lupus_familiaris_Dalmatian                              GGGGTAGGAATGACAAGAATTATAAGAGTCAAATATAGAGGCTGTGGGCGAGGCTTCAATGATAGGTCACATCTCAGC
JF750644.1|Canis_lupus_familiaris_Labrador_retriever                     GGGGTAGGAATGACAAGAATTATAAGAGTCAAATATAGAGGCTGTGGGCGAGGCTTCAATGATAGGTCACATCTCAGC
JF750645.1|Canis_lupus_familiaris_German_Shepard                         GGGGTAGGAATGACAAGAATTATAAGAGTCAAATATAGAGGCTGTGGGCGAGGCTTCAATGATAGGTCACATCTCAGC
JF750646.1|Canis_lupus_familiaris_Rough_Collie                           GGGGTAGGAATGACAAGAATTATAAGAGTCAAATATAGAGGCTGTGGGCGAGGCTTCAATGATAGGTCACATCTCAGC
JF750647.1|Canis_lupus_familiaris_Eurasier                               GGGGTAGGAATGACAAGAATTATAAGAGTCAAATATAGAGGCTGTGGGCGAGGCTTCAATGATAGGTCACATCTCAGC
JF750648.1|Canis_lupus_familiaris_Polish_Lowland_Sheepdog                GGGGTAGGAATGACAAGAATTATAAGAGTCAAATATAGAGGCTGTGGGCGAGGCTTCAATGATAGGTCACATCTCAGC
JF750649.1|Canis_lupus_familiaris_Rottweiler                             GGGGTAGGAATGACAAGAATTATAAGAGTCAAATATAGAGGCTGTGGGCGAGGCTTCAATGATAGGTCACATCTCAGC
JF750650.1|Canis_lupus_familiaris_Smalandsstrovare                       GGGGTAGGAATGACAAGAATTATAAGAGTCAAATATAGAGGCTGTGGGCGAGGCTTCAATGATAGGTCACATCTCAGC
JF750651.1|Canis_lupus_familiaris_Swedish_Elkhound                       GGGGTAGGAATGACAAGAATTATAAGAGTCAAATATAGAGGCTGTGGGCGAGGCTTCAATGATAGGTCACATCTCAGC
JF750652.1|Canis_lupus_familiaris_Nova_Scotia_Duck_Tolling_Retriever     GGGGTAGGAATGACAAGAATTATAAGAGTCAAATATAGAGGCTGTGGGCGAGGCTTCAATGATAGGTCACATCTCAGC
JF750653.1|Canis_lupus_isolate_ref21                                     GGGGTAGGAATGACAAGAATTATAAGAGTCAAATATAGAGGCTGTGGGCGAGGCTTCAATGATAGGTCACATCTCAGC
JF750654.1|Canis_lupus_isolate_ref34                                     GGGGTAGGAATGACAAGAATTATAAGAGTCAAATATAGAGGCTGTGGGCGAGGCTTCAATGATAGGTCACATCTCAGC
JF750655.1|Vulpes_vulpes                                                 GGGGTAGGAATGACAAGAATTATAAGAGTCAAATATAGAGGCTGTGGGCGAGGCTTCAATGATAGGTCACATCTCAGC
JF750656.1|Speothos_venaticus                                            GGGGTAGGAATGACAAGAATTATAAGAGTCAAATATAGAGGCTGTGGGCGAGGCTTCAATGATAGGTCACATCTCAGC
JF750657.1|Lycaon_pictus                                                 GGGGTAGGAATGACAAGAATTAT--GAGTCAAATATAGAGGCTGTGGGCGAGGCTTCAATGATAGGTCACATCTCAGC
JF750658.1|Canis_mesomelas                                               GGGGTAGGAATGACAAGAATTATAAGAGTCAAATATAGAGGCTGTGGGCGAGGCTTCAATGATAGGTCACATCTCAGC
JF750659.1|Canis_aureus                                                  GGGGTAGGAATGACAAGAATTATAAGAGTCAAATATAGAGGCTGTGGGCGAGGCTTCAATGATAGGTCACATCTCAGC
Island_Fox|Urocyon littoralis                                            GGGGTAGGAATGACAAGAATTATAAGAGTCAAATATAGAGGCTGTGGGCAAGGCTTCAATGATAGGTCACATCTCAGC
Culpeo|Lycalopex culpaeus                                                GGGGTAGGAATGACAAGAATTATAAGAGTCAAATATAGAGGCTGTGGGCGAGGCTTCAATGATAGGTCACATCTCAGC
```

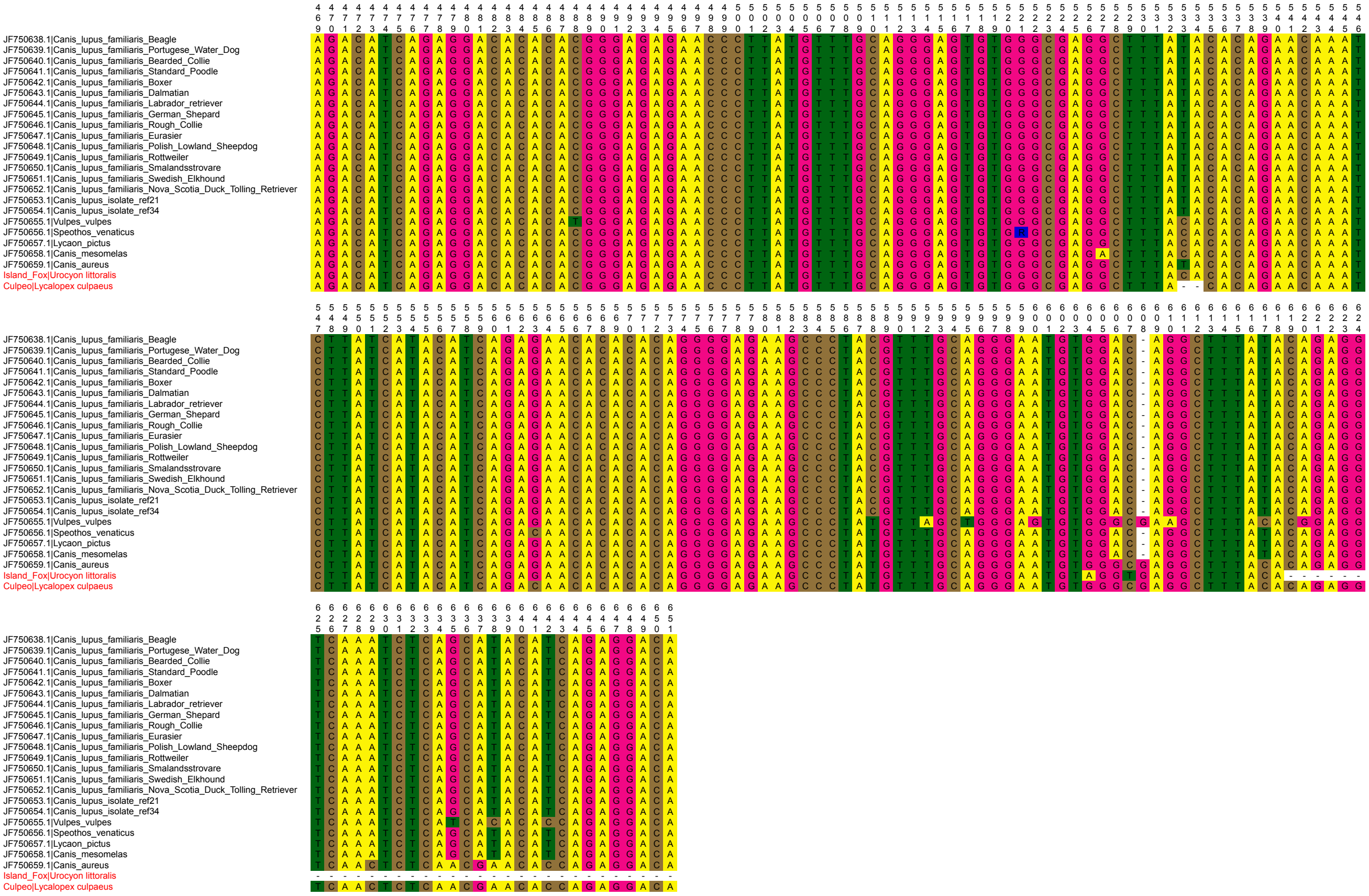
JF750638.1|Canis_lupus_familiaris_Beagle
JF750639.1|Canis_lupus_familiaris_Portugese_Water_Dog
JF750640.1|Canis_lupus_familiaris_Bearded_Collie
JF750641.1|Canis_lupus_familiaris_Standard_Poodle
JF750642.1|Canis_lupus_familiaris_Boxer
JF750643.1|Canis_lupus_familiaris_Dalmatian
JF750644.1|Canis_lupus_familiaris_Labrador_retriever
JF750645.1|Canis_lupus_familiaris_German_Shepard
JF750646.1|Canis_lupus_familiaris_Rough_Collie
JF750647.1|Canis_lupus_familiaris_Eurasier
JF750648.1|Canis_lupus_familiaris_Polish_Lowland_Sheepdog
JF750649.1|Canis_lupus_familiaris_Rottweiler
JF750650.1|Canis_lupus_familiaris_Smalandsstrovare
JF750651.1|Canis_lupus_familiaris_Swedish_Elkhound
JF750652.1|Canis_lupus_familiaris_Nova_Scotia_Duck_Tolling_Retriever
JF750653.1|Canis_lupus_isolate_ref21
JF750654.1|Canis_lupus_isolate_ref34
JF750655.1|Vulpes_vulpes
JF750656.1|Speothos_venaticus
JF750657.1|Lycaon_pictus
JF750658.1|Canis_mesomelas
JF750659.1|Canis_aureus
Island_Fox|Urocyon littoralis
Culpeo|Lycalopex culpaeus

**Supplementary Table 5: PRDM9 exon 7 Amino Acid sequences**

|  | 1 | 2 | 3 | 4 | 5 | 6 | 7 | 8 | 9 | 10 | 11 | 12 | 13 | 14 | 15 | 16 | 17 | 18 | 19 | 20 | 21 | 22 | 23 | 24 | 25 | 26 | 27 | 28 | 29 | 30 | 31 | 32 | 33 | 34 | 35 | 36 | 37 | 38 | 39 | 40 | 41 | 42 | 43 | 44 | 45 | 46 | 47 | 48 | 49 | 50 | 51 | 52 | 53 | 54 | 55 | 56 | 57 | 58 | 59 | 60 | 61 | 62 | 63 | 64 | 65 | 66 | 67 | 68 | 69 | 70 | 71 | 72 | 73 | 74 | 75 | 76 | 77 | 78 | 79 | 80 |
|---|---|---|---|---|---|---|---|---|---|---|---|---|---|---|---|---|---|---|---|---|---|---|---|---|---|---|---|---|---|---|---|---|---|---|---|---|---|---|---|---|---|---|---|---|---|---|---|---|---|---|---|---|---|---|---|---|---|---|---|---|---|---|---|---|---|---|---|---|---|---|---|---|---|---|---|---|---|---|---|
| JF750638.1\|Canis_lupus_familiaris_Beagle | P | N | P | E | I | H | P | C | P | S | C | S | L | - | F | S | S | Q | K | F | L | S | Q | H | L | E | H | ? | H | P | S | Q | I | L | P | R | I | S | V | R | E | H | F | R | P | K | D | P | C | P | G | C | Q | N | Q | Q | Q | Q | Q | H | S | D | P | Q | R | W | N | D | R | A | K | G | Q | E | G | K | E | R | F | K |
| JF750639.1\|Canis_lupus_familiaris_Portugese_Water_Dog | P | N | P | E | I | H | P | C | P | S | C | S | L | - | F | S | S | Q | K | F | L | S | Q | H | L | E | H | N | H | P | S | Q | I | L | P | R | I | S | V | R | E | H | F | R | P | K | D | P | C | P | G | C | Q | N | Q | Q | Q | Q | Q | H | S | D | P | Q | R | W | N | D | R | A | K | G | Q | E | G | K | E | R | F | K |
| JF750640.1\|Canis_lupus_familiaris_Bearded_Collie | P | N | P | E | I | H | P | C | P | S | C | S | L | - | F | S | S | Q | K | F | L | S | Q | H | L | E | H | N | H | P | S | Q | I | L | P | R | I | S | V | R | E | H | F | R | P | K | D | P | C | P | G | C | Q | N | Q | Q | Q | Q | Q | H | S | D | P | Q | R | W | N | D | R | A | K | G | Q | E | G | K | E | R | F | K |
| JF750641.1\|Canis_lupus_familiaris_Standard_Poodle | P | N | P | E | I | H | P | C | P | S | C | S | L | - | F | S | S | Q | K | F | L | S | Q | H | L | E | H | N | H | P | S | Q | I | L | P | R | I | S | V | R | E | H | F | R | P | K | D | P | C | P | G | C | Q | N | Q | Q | Q | Q | Q | H | S | D | P | Q | R | W | N | D | R | A | K | G | Q | E | G | K | E | R | F | K |
| JF750642.1\|Canis_lupus_familiaris_Boxer | P | N | P | E | I | H | P | C | P | S | C | S | L | - | F | S | S | Q | K | F | L | S | Q | H | L | E | H | N | H | P | S | Q | I | L | P | R | I | S | V | R | E | H | F | R | P | K | D | P | C | P | G | C | Q | N | Q | Q | Q | Q | Q | H | S | D | P | Q | R | W | N | D | R | A | K | G | Q | E | G | K | E | R | F | K |
| JF750643.1\|Canis_lupus_familiaris_Dalmatian | P | N | P | E | I | H | P | C | P | S | C | S | L | - | F | S | S | Q | K | F | L | S | Q | H | L | E | H | N | H | P | S | Q | I | L | P | R | I | S | V | R | E | H | F | R | P | K | D | P | C | P | G | C | Q | N | Q | Q | Q | Q | Q | H | S | D | P | Q | R | W | N | D | R | A | K | G | Q | E | G | K | E | R | F | K |
| JF750644.1\|Canis_lupus_familiaris_Labrador_retriever | P | N | P | E | I | H | P | C | P | S | C | S | L | - | F | S | S | Q | K | F | L | S | Q | H | L | E | H | N | H | P | S | Q | I | L | P | R | I | S | V | R | E | H | F | R | P | K | D | P | C | P | G | C | Q | N | Q | Q | Q | Q | Q | H | S | D | P | Q | R | W | N | D | R | A | K | G | Q | E | G | K | E | R | F | K |
| JF750645.1\|Canis_lupus_familiaris_German_Shepard | P | N | P | E | I | H | P | C | P | S | C | S | L | - | F | S | S | Q | K | F | L | S | Q | H | L | E | H | N | H | P | S | Q | I | L | P | R | I | S | V | R | E | H | F | R | P | K | D | P | C | P | G | C | Q | N | Q | Q | Q | Q | Q | H | S | D | P | Q | R | W | N | D | R | A | K | G | Q | E | G | K | E | R | F | K |
| JF750646.1\|Canis_lupus_familiaris_Rough_Collie | P | N | P | E | I | H | P | C | P | S | C | S | L | - | F | S | S | Q | K | F | L | S | Q | H | L | E | H | N | H | P | S | Q | I | L | P | R | I | S | V | R | E | H | F | R | P | K | D | P | C | P | G | C | Q | N | Q | Q | Q | Q | Q | H | S | D | P | Q | R | W | N | D | R | A | K | G | Q | E | G | K | E | R | F | K |
| JF750647.1\|Canis_lupus_familiaris_Eurasier | P | N | P | E | I | H | P | C | P | S | C | S | L | - | F | S | S | Q | K | F | L | S | Q | H | L | E | H | N | H | P | S | Q | I | L | P | R | I | S | V | R | E | H | F | R | P | K | D | P | C | P | G | C | Q | N | Q | Q | Q | Q | Q | H | S | D | P | Q | R | W | N | D | R | A | K | G | Q | E | G | K | E | R | F | K |
| JF750648.1\|Canis_lupus_familiaris_Polish_Lowland_Sheepdog | P | N | P | E | I | ? | ? | ? | P | S | C | S | L | - | F | S | S | Q | K | F | L | S | Q | H | L | E | H | N | H | P | S | Q | I | L | P | R | I | S | V | R | E | H | F | R | P | K | D | P | C | P | G | C | Q | N | Q | Q | Q | Q | Q | H | S | D | P | Q | R | W | N | D | R | A | K | G | Q | E | G | K | E | R | F | K |
| JF750649.1\|Canis_lupus_familiaris_Rottweiler | P | N | P | E | I | H | P | C | P | S | C | S | L | - | F | S | S | Q | K | F | L | S | Q | H | L | E | H | ? | H | P | S | Q | I | L | P | R | I | S | V | R | E | H | F | R | P | K | D | P | C | P | G | C | Q | N | Q | Q | Q | Q | Q | H | S | D | P | Q | R | W | N | D | R | A | K | G | Q | E | G | K | E | R | F | K |
| JF750650.1\|Canis_lupus_familiaris_Smalandsstrovare | P | N | P | E | I | H | P | C | P | S | C | S | L | - | F | S | S | Q | K | F | L | S | ? | H | L | E | H | N | H | P | S | Q | I | L | P | R | I | S | V | R | E | H | F | R | P | K | D | P | C | P | G | C | Q | N | Q | Q | Q | Q | Q | H | S | D | P | Q | R | W | N | D | R | A | K | G | Q | E | G | K | E | R | F | K |
| JF750651.1\|Canis_lupus_familiaris_Swedish_Elkhound | P | N | P | E | I | ? | ? | ? | P | S | C | S | L | - | F | S | S | Q | K | F | L | S | Q | H | L | E | H | ? | H | P | S | Q | I | L | P | R | I | S | V | R | E | H | F | R | P | K | D | P | C | P | G | C | Q | N | Q | Q | Q | Q | Q | H | S | D | P | Q | R | W | N | D | R | A | K | G | Q | E | G | K | E | R | F | K |
| JF750652.1\|Canis_lupus_familiaris_Nova_Scotia_Duck_Tolling_Retriever | P | N | P | E | I | H | P | C | P | S | C | S | L | - | F | S | S | Q | K | F | L | S | Q | H | L | E | H | N | H | P | S | Q | I | L | P | R | I | S | V | R | E | H | F | R | P | K | D | P | C | P | G | C | Q | N | Q | Q | Q | Q | Q | H | S | D | P | Q | R | W | N | D | R | A | K | G | Q | E | G | K | E | R | F | K |
| JF750653.1\|Canis_lupus_isolate_ref21 | P | N | P | E | I | H | P | C | P | S | C | S | L | - | F | S | S | Q | K | F | L | S | Q | H | L | E | H | N | H | P | S | Q | I | L | P | R | I | S | V | R | E | H | F | R | P | K | D | P | C | P | G | C | Q | N | Q | Q | Q | Q | Q | H | S | D | P | Q | R | W | N | D | R | A | K | G | Q | E | G | K | E | R | F | K |
| JF750654.1\|Canis_lupus_isolate_ref34 | P | N | P | E | I | H | P | C | P | S | C | S | L | - | F | S | S | Q | K | F | L | S | Q | H | L | E | H | N | H | P | S | Q | I | L | P | R | I | S | V | R | E | H | F | R | P | K | D | P | C | P | G | C | Q | N | Q | Q | Q | Q | Q | H | S | D | P | Q | R | W | N | D | R | A | K | G | Q | E | G | K | E | R | F | K |
| JF750655.1\|Vulpes_vulpes | P | N | P | E | I | Y | P | C | P | S | C | S | L | S | F | S | S | Q | K | F | L | S | Q | H | L | E | H | N | H | P | S | Q | I | L | P | R | I | S | I | R | E | H | F | Q | P | K | D | P | C | P | G | C | Q | N | Q | Q | Q | Q | Q | H | S | D | P | Q | C | W | N | D | R | A | K | G | Q | E | G | K | E | R | F | K |
| JF750656.1\|Speothos_venaticus | P | N | P | E | I | H | P | C | P | S | C | S | L | - | F | S | S | Q | K | F | L | S | Q | H | L | E | H | N | H | P | S | Q | I | L | P | * | I | S | V | R | E | H | F | R | P | K | D | P | C | P | G | C | Q | N | Q | Q | Q | Q | Q | H | S | D | P | Q | R | W | N | D | R | A | K | G | Q | E | G | K | E | R | F | K |
| JF750657.1\|Lycaon_pictus | P | N | P | E | I | H | P | C | P | S | C | S | L | - | F | S | S | Q | K | F | L | S | Q | H | L | E | H | N | H | P | S | Q | I | L | P | R | I | S | V | R | E | H | F | R | P | K | D | P | C | P | G | C | Q | N | Q | Q | Q | Q | Q | H | S | D | P | Q | R | W | S | D | R | A | K | G | Q | E | G | K | E | R | F | K |
| JF750658.1\|Canis_mesomelas | P | N | P | E | I | H | P | C | P | S | C | S | L | - | F | S | S | Q | K | F | L | S | Q | H | L | E | H | N | H | P | S | Q | I | L | P | Q | I | S | V | R | E | H | F | R | P | K | D | P | C | P | G | C | Q | N | Q | Q | Q | Q | Q | H | S | D | P | Q | R | W | N | D | R | A | K | G | Q | E | G | K | E | R | F | K |
| JF750659.1\|Canis_aureus | P | N | P | E | I | H | P | C | P | S | C | S | L | - | F | S | S | Q | K | F | L | S | Q | H | L | E | H | N | H | P | S | Q | I | L | P | R | I | S | V | R | E | H | F | R | P | K | D | P | C | P | G | C | Q | N | Q | Q | Q | Q | Q | H | S | D | P | Q | R | W | N | D | R | A | K | G | Q | E | G | K | E | R | F | K |
| Island_Fox\|Urocyon littoralis | ? | N | P | E | I | H | P | C | P | S | C | S | L | A | F | S | S | Q | K | F | L | S | Q | H | L | E | H | N | H | P | ? | Q | I | L | P | R | I | S | V | R | E | H | F | R | P | K | D | P | C | P | G | C | Q | N | Q | Q | Q | Q | Q | H | S | D | P | Q | R | W | N | D | R | A | K | G | Q | E | G | K | E | R | F | K |
| Culpeo\|Lycalopex culpaeus | P | N | P | E | I | H | P | C | P | S | C | S | L | - | F | S | S | Q | K | F | L | S | Q | H | L | E | H | N | H | P | S | Q | I | L | P | R | I | S | V | R | E | H | F | * | P | K | D | P | C | P | G | C | Q | N | Q | Q | Q | Q | Q | H | S | D | P | Q | R | W | N | D | R | A | K | G | Q | E | G | K | E | R | F | K |

|  | 81 | 82 | 83 | 84 | 85 | 86 | 87 | 88 | 89 | 90 | 91 | 92 | 93 | 94 | 95 | 96 | 97 | 98 | 99 | 100 | 101 | 102 | 103 | 104 | 105 | 106 | 107 | 108 | 109 | 110 | 111 | 112 | 113 | 114 | 115 | 116 | 117 | 118 | 119 | 120 | 121 | 122 | 123 | 124 | 125 | 126 | 127 | 128 | 129 | 130 | 131 | 132 | 133 | 134 | 135 | 136 | 137 | 138 | 139 | 140 | 141 | 142 | 143 | 144 | 145 | 146 | 147 | 148 | 149 | 150 | 151 | 152 | 153 | 154 | 155 | 156 | 157 | 158 | 159 | 160 |
|---|---|---|---|---|---|---|---|---|---|---|---|---|---|---|---|---|---|---|---|---|---|---|---|---|---|---|---|---|---|---|---|---|---|---|---|---|---|---|---|---|---|---|---|---|---|---|---|---|---|---|---|---|---|---|---|---|---|---|---|---|---|---|---|---|---|---|---|---|---|---|---|---|---|---|---|---|---|---|---|
| JF750638.1\|Canis_lupus_familiaris_Beagle | P | L | P | K | S | I | R | Q | R | R | I | S | R | A | F | S | T | P | C | K | G | Q | T | T | C | E | G | I | V | K | E | E | P | S | A | G | S | Q | K | L | N | P | E | D | T | G | K | L | F | K | G | V | G | M | T | R | I | I | R | V | K | Y | R | G | C | G | R | G | F | N | D | R | S | H | L | S | R | H | Q | R |
| JF750639.1\|Canis_lupus_familiaris_Portugese_Water_Dog | P | L | P | K | S | I | R | Q | R | R | I | S | R | A | F | S | T | P | C | K | G | Q | T | T | C | E | G | I | V | K | E | E | P | S | A | G | S | Q | K | L | N | P | E | D | T | G | K | L | F | K | G | V | G | M | T | R | I | I | R | V | K | Y | R | G | C | G | R | G | F | N | D | R | S | H | L | S | R | H | Q | R |
| JF750640.1\|Canis_lupus_familiaris_Bearded_Collie | P | L | P | K | S | I | R | Q | R | R | I | S | R | A | F | S | T | P | C | K | G | Q | T | T | C | E | G | I | V | K | E | E | P | S | A | G | S | Q | K | L | N | P | E | D | T | G | K | L | F | K | G | V | G | M | T | R | I | I | R | V | K | Y | R | G | C | G | R | G | F | N | D | R | S | H | L | S | R | H | Q | R |
| JF750641.1\|Canis_lupus_familiaris_Standard_Poodle | P | L | P | K | S | I | R | Q | R | R | I | S | R | A | F | S | T | P | C | K | G | Q | T | T | C | E | G | I | V | K | E | E | P | S | A | G | S | Q | K | L | N | P | E | D | T | G | K | L | F | K | G | V | G | M | T | R | I | I | R | V | K | Y | R | G | C | G | R | G | F | N | D | R | S | H | L | S | R | H | Q | R |
| JF750642.1\|Canis_lupus_familiaris_Boxer | P | L | P | K | S | I | R | Q | R | R | I | S | R | A | F | S | T | P | C | K | G | Q | T | T | C | E | G | I | V | K | E | E | P | S | A | G | S | Q | K | L | N | P | E | D | T | G | K | L | F | K | G | V | G | M | T | R | I | I | R | V | K | Y | R | G | C | G | R | G | F | N | D | R | S | H | L | S | R | H | Q | R |
| JF750643.1\|Canis_lupus_familiaris_Dalmatian | P | L | P | K | S | I | R | Q | R | R | I | S | R | A | F | S | T | P | C | K | G | Q | T | T | C | E | G | I | V | K | E | E | P | S | A | G | S | Q | K | L | N | P | E | D | T | G | K | L | F | K | G | V | G | M | T | R | I | I | R | V | K | Y | R | G | C | G | R | G | F | N | D | R | S | H | L | S | R | H | Q | R |
| JF750644.1\|Canis_lupus_familiaris_Labrador_retriever | P | L | P | K | S | I | R | Q | R | R | I | S | R | A | F | S | T | P | C | K | G | Q | T | T | C | E | G | I | V | K | E | E | P | S | A | G | S | Q | K | L | N | P | E | D | T | G | K | L | F | K | G | V | G | M | T | R | I | I | R | V | K | Y | R | G | C | G | R | G | F | N | D | R | S | H | L | S | R | H | Q | R |
| JF750645.1\|Canis_lupus_familiaris_German_Shepard | P | L | P | K | S | I | R | Q | R | R | I | S | R | A | F | S | T | P | C | K | G | Q | T | T | C | E | G | I | V | K | E | E | P | S | A | G | S | Q | K | L | N | P | E | D | T | G | K | L | F | K | G | V | G | M | T | R | I | I | R | V | K | Y | R | G | C | G | R | G | F | N | D | R | S | H | L | S | R | H | Q | R |
| JF750646.1\|Canis_lupus_familiaris_Rough_Collie | P | L | P | K | S | I | R | Q | R | R | I | S | R | A | F | S | T | P | C | K | G | Q | T | T | C | E | G | I | V | K | E | E | P | S | A | G | S | Q | K | L | N | P | E | D | T | G | K | L | F | K | G | V | G | M | T | R | I | I | R | V | K | Y | R | G | C | G | R | G | F | N | D | R | S | H | L | S | R | H | Q | R |
| JF750647.1\|Canis_lupus_familiaris_Eurasier | P | L | P | K | S | I | R | Q | R | R | I | S | R | A | F | S | T | P | C | K | G | Q | T | T | C | E | G | I | V | K | E | E | P | S | A | G | S | Q | K | L | N | P | E | D | T | G | K | L | F | K | G | V | G | M | T | R | I | I | R | V | K | Y | R | G | C | G | R | G | F | N | D | R | S | H | L | S | R | H | Q | R |
| JF750648.1\|Canis_lupus_familiaris_Polish_Lowland_Sheepdog | P | L | P | K | S | I | R | Q | R | R | I | S | R | A | F | S | T | P | C | K | G | Q | T | T | C | E | G | I | V | K | E | E | P | S | A | G | S | Q | K | L | N | P | E | D | T | G | K | L | F | K | G | V | G | M | T | R | I | I | R | V | K | Y | R | G | C | G | R | G | F | N | D | R | S | H | L | S | R | H | Q | R |
| JF750649.1\|Canis_lupus_familiaris_Rottweiler | P | L | P | K | S | I | R | Q | R | R | I | S | R | A | F | S | T | P | C | K | G | Q | T | T | C | E | G | I | V | K | E | E | P | S | A | G | S | Q | K | L | N | P | E | D | T | G | K | L | F | K | G | V | G | M | T | R | I | I | R | V | K | Y | R | G | C | G | R | G | F | N | D | R | S | H | L | S | R | H | Q | R |
| JF750650.1\|Canis_lupus_familiaris_Smalandsstrovare | P | L | P | K | S | I | R | Q | R | R | I | S | R | A | F | S | T | P | C | K | G | Q | T | T | C | E | G | I | V | K | E | E | P | S | A | G | S | Q | K | L | N | P | E | D | T | G | K | L | F | K | G | V | G | M | T | R | I | I | R | V | K | Y | R | G | C | G | R | G | F | N | D | R | S | H | L | S | R | H | Q | R |
| JF750651.1\|Canis_lupus_familiaris_Swedish_Elkhound | P | L | P | K | S | I | R | Q | R | R | I | S | R | A | F | S | T | P | C | K | G | Q | T | T | C | E | G | I | V | K | E | E | P | S | A | G | S | Q | K | L | N | P | E | D | T | G | K | L | F | K | G | V | G | M | T | R | I | I | R | V | K | Y | R | G | C | G | R | G | F | N | D | R | S | H | L | S | R | H | Q | R |
| JF750652.1\|Canis_lupus_familiaris_Nova_Scotia_Duck_Tolling_Retriever | P | L | P | K | S | I | R | Q | R | R | I | S | R | A | F | S | T | P | C | K | G | Q | T | T | C | E | G | I | V | K | E | E | P | S | A | G | S | Q | K | L | N | P | E | D | T | G | K | L | F | K | G | V | G | M | T | R | I | I | R | V | K | Y | R | G | C | G | R | G | F | N | D | R | S | H | L | S | R | H | Q | R |
| JF750653.1\|Canis_lupus_isolate_ref21 | P | L | P | K | S | I | R | Q | R | R | I | S | R | A | F | S | T | P | C | K | G | Q | T | T | C | E | G | I | V | K | E | E | P | S | A | G | S | Q | K | L | N | P | E | D | T | G | K | L | F | K | G | V | G | M | T | R | I | I | R | V | K | Y | R | G | C | G | R | G | F | N | D | R | S | H | L | S | R | H | Q | R |
| JF750654.1\|Canis_lupus_isolate_ref34 | P | L | P | K | S | I | R | Q | R | R | I | S | R | A | F | S | T | P | C | K | G | Q | T | T | C | E | G | I | V | K | E | E | P | S | A | G | S | Q | K | L | N | P | E | D | T | G | K | L | F | K | G | V | G | M | T | R | I | I | R | V | K | Y | R | G | C | G | R | G | F | N | D | R | S | H | L | S | R | H | Q | R |
| JF750655.1\|Vulpes_vulpes | P | L | P | K | - | I | R | Q | R | K | I | S | R | A | F | S | T | P | C | K | G | Q | T | T | C | E | G | I | V | K | E | E | P | S | A | G | S | Q | K | L | N | P | E | D | T | G | K | L | F | K | G | V | G | M | T | R | I | I | R | V | K | Y | R | G | C | G | R | G | F | N | D | R | S | H | L | S | R | H | Q | R |
| JF750656.1\|Speothos_venaticus | P | L | P | K | R | I | R | Q | R | R | I | S | R | A | F | S | T | P | C | K | G | Q | T | T | C | E | G | I | V | K | E | E | P | S | A | S | S | Q | K | L | N | P | E | ? | T | G | K | L | F | K | G | V | G | M | T | R | I | I | R | V | K | Y | R | G | C | G | R | G | F | N | D | R | S | H | L | S | R | H | Q | R |
| JF750657.1\|Lycaon_pictus | P | L | P | K | S | I | R | Q | R | R | I | S | R | A | F | S | T | P | C | K | G | Q | T | T | C | E | G | I | V | K | E | E | P | S | A | G | S | Q | K | L | N | P | E | D | T | G | K | L | F | K | G | V | G | M | T | R | I | - | - | V | K | Y | R | G | C | G | R | G | F | N | D | R | S | H | L | S | R | H | Q | R |
| JF750658.1\|Canis_mesomelas | P | L | P | K | S | I | R | Q | R | R | I | S | R | A | F | S | T | P | C | K | G | Q | T | T | C | E | G | I | V | K | E | E | P | S | A | G | S | Q | K | L | N | P | E | D | T | G | K | L | F | K | G | V | G | M | T | R | I | I | R | V | K | Y | R | G | C | G | R | G | F | N | D | R | S | H | L | S | R | H | Q | R |
| JF750659.1\|Canis_aureus | P | L | P | K | S | I | R | Q | R | R | I | S | R | A | F | S | T | P | C | K | G | Q | T | T | C | E | G | I | V | K | E | E | P | S | A | G | S | Q | K | L | N | P | E | D | T | G | K | L | F | K | G | V | G | M | T | R | I | I | R | V | K | Y | R | G | C | G | R | G | F | N | D | R | S | H | L | S | R | H | Q | R |
| Island_Fox\|Urocyon littoralis | P | L | P | K | S | I | R | Q | R | R | I | S | R | A | F | S | T | P | C | K | G | Q | T | T | C | E | G | I | V | K | E | E | P | S | T | G | S | Q | K | L | N | P | E | D | T | G | K | L | F | K | G | V | G | M | T | R | I | I | R | V | K | Y | R | G | C | G | Q | G | F | N | D | R | S | H | L | S | R | H | Q | R |
| Culpeo\|Lycalopex culpaeus | P | L | P | K | R | I | R | Q | R | R | I | S | R | A | F | S | T | P | C | K | G | Q | T | T | C | E | G | I | V | K | E | E | P | S | A | G | S | Q | K | L | N | P | E | D | T | G | K | L | F | K | G | V | G | M | T | R | I | I | R | V | K | Y | R | G | C | G | R | G | F | N | D | R | S | H | L | S | R | H | Q | R |

|  | 161 | 162 | 163 | 164 | 165 | 166 | 167 | 168 | 169 | 170 | 171 | 172 | 173 | 174 | 175 | 176 | 177 | 178 | 179 | 180 | 181 | 182 | 183 | 184 | 185 | 186 | 187 | 188 | 189 | 190 | 191 | 192 | 193 | 194 | 195 | 196 | 197 | 198 | 199 | 200 | 201 | 202 | 203 | 204 | 205 | 206 | 207 | 208 | 209 | 210 | 211 | 212 | 213 | 214 | 215 | 216 | 217 |
|---|---|---|---|---|---|---|---|---|---|---|---|---|---|---|---|---|---|---|---|---|---|---|---|---|---|---|---|---|---|---|---|---|---|---|---|---|---|---|---|---|---|---|---|---|---|---|---|---|---|---|---|---|---|---|---|---|
| JF750638.1\|Canis_lupus_familiaris_Beagle | T | H | T | G | E | N | P | Y | V | C | R | E | C | G | R | G | F | I | H | R | T | N | L | I | I | H | Q | R | T | H | T | G | E | K | P | Y | V | C | R | E | C | G | - | G | F | I | Q | R | S | N | L | S | I | H | Q | R | T |
| JF750639.1\|Canis_lupus_familiaris_Portugese_Water_Dog | T | H | T | G | E | N | P | Y | V | C | R | E | C | G | R | G | F | I | H | R | T | N | L | I | I | H | Q | R | T | H | T | G | E | K | P | Y | V | C | R | E | C | G | - | G | F | I | Q | R | S | N | L | S | I | H | Q | R | T |
| JF750640.1\|Canis_lupus_familiaris_Bearded_Collie | T | H | T | G | E | N | P | Y | V | C | R | E | C | G | R | G | F | I | H | R | T | N | L | I | I | H | Q | R | T | H | T | G | E | K | P | Y | V | C | R | E | C | G | - | G | F | I | Q | R | S | N | L | S | I | H | Q | R | T |
| JF750641.1\|Canis_lupus_familiaris_Standard_Poodle | T | H | T | G | E | N | P | Y | V | C | R | E | C | G | R | G | F | I | H | R | T | N | L | I | I | H | Q | R | T | H | T | G | E | K | P | Y | V | C | R | E | C | G | - | G | F | I | Q | R | S | N | L | S | I | H | Q | R | T |
| JF750642.1\|Canis_lupus_familiaris_Boxer | T | H | T | G | E | N | P | Y | V | C | R | E | C | G | R | G | F | I | H | R | T | N | L | I | I | H | Q | R | T | H | T | G | E | K | P | Y | V | C | R | E | C | G | - | G | F | I | Q | R | S | N | L | S | I | H | Q | R | T |
| JF750643.1\|Canis_lupus_familiaris_Dalmatian | T | H | T | G | E | N | P | Y | V | C | R | E | C | G | R | G | F | I | H | R | T | N | L | I | I | H | Q | R | T | H | T | G | E | K | P | Y | V | C | R | E | C | G | - | G | F | I | Q | R | S | N | L | S | I | H | Q | R | T |
| JF750644.1\|Canis_lupus_familiaris_Labrador_retriever | T | H | T | G | E | N | P | Y | V | C | R | E | C | G | R | G | F | I | H | R | T | N | L | I | I | H | Q | R | T | H | T | G | E | K | P | Y | V | C | R | E | C | G | - | G | F | I | Q | R | S | N | L | S | I | H | Q | R | T |
| JF750645.1\|Canis_lupus_familiaris_German_Shepard | T | H | T | G | E | N | P | Y | V | C | R | E | C | G | R | G | F | I | H | R | T | N | L | I | I | H | Q | R | T | H | T | G | E | K | P | Y | V | C | R | E | C | G | - | G | F | I | Q | R | S | N | L | S | I | H | Q | R | T |
| JF750646.1\|Canis_lupus_familiaris_Rough_Collie | T | H | T | G | E | N | P | Y | V | C | R | E | C | G | R | G | F | I | H | R | T | N | L | I | I | H | Q | R | T | H | T | G | E | K | P | Y | V | C | R | E | C | G | - | G | F | I | Q | R | S | N | L | S | I | H | Q | R | T |
| JF750647.1\|Canis_lupus_familiaris_Eurasier | T | H | T | G | E | N | P | Y | V | C | R | E | C | G | R | G | F | I | H | R | T | N | L | I | I | H | Q | R | T | H | T | G | E | K | P | Y | V | C | R | E | C | G | - | G | F | I | Q | R | S | N | L | S | I | H | Q | R | T |
| JF750648.1\|Canis_lupus_familiaris_Polish_Lowland_Sheepdog | T | H | T | G | E | N | P | Y | V | C | R | E | C | G | R | G | F | I | H | R | T | N | L | I | I | H | Q | R | T | H | T | G | E | K | P | Y | V | C | R | E | C | G | - | G | F | I | Q | R | S | N | L | S | I | H | Q | R | T |
| JF750649.1\|Canis_lupus_familiaris_Rottweiler | T | H | T | G | E | N | P | Y | V | C | R | E | C | G | R | G | F | I | H | R | T | N | L | I | I | H | Q | R | T | H | T | G | E | K | P | Y | V | C | R | E | C | G | - | G | F | I | Q | R | S | N | L | S | I | H | Q | R | T |
| JF750650.1\|Canis_lupus_familiaris_Smalandsstrovare | T | H | T | G | E | N | P | Y | V | C | R | E | C | G | R | G | F | I | H | R | T | N | L | I | I | H | Q | R | T | H | T | G | E | K | P | Y | V | C | R | E | C | G | - | G | F | I | Q | R | S | N | L | S | I | H | Q | R | T |
| JF750651.1\|Canis_lupus_familiaris_Swedish_Elkhound | T | H | T | G | E | N | P | Y | V | C | R | E | C | G | R | G | F | I | H | R | T | N | L | I | I | H | Q | R | T | H | T | G | E | K | P | Y | V | C | R | E | C | G | - | G | F | I | Q | R | S | N | L | S | I | H | Q | R | T |
| JF750652.1\|Canis_lupus_familiaris_Nova_Scotia_Duck_Tolling_Retriever | T | H | T | G | E | N | P | Y | V | C | R | E | C | G | R | G | F | I | H | R | T | N | L | I | I | H | Q | R | T | H | T | G | E | K | P | Y | V | C | R | E | C | G | - | G | F | I | Q | R | S | N | L | S | I | H | Q | R | T |
| JF750653.1\|Canis_lupus_isolate_ref21 | T | H | T | G | E | N | P | Y | V | C | R | E | C | G | R | G | F | I | H | R | T | N | L | I | I | H | Q | R | T | H | T | G | E | K | P | Y | V | C | R | E | C | G | - | G | F | I | Q | R | S | N | L | S | I | H | Q | R | T |
| JF750654.1\|Canis_lupus_isolate_ref34 | T | H | T | G | E | N | P | Y | V | C | R | E | C | G | R | G | F | I | H | R | T | N | L | I | I | H | Q | R | T | H | T | G | E | K | P | Y | V | C | R | E | C | G | - | G | F | I | Q | R | S | N | L | S | I | H | Q | R | T |
| JF750655.1\|Vulpes_vulpes | T | H | M | G | E | N | P | Y | V | C | R | E | C | G | R | G | F | T | H | R | T | N | L | I | I | H | Q | R | T | H | T | G | E | K | P | Y | V | S | W | E | C | G | R | S | F | T | R | R | S | N | L | I | T | H | Q | R | T |
| JF750656.1\|Speothos_venaticus | T | H | T | G | E | N | P | Y | V | C | R | E | C | ? | R | G | F | T | H | R | T | N | L | I | I | H | Q | T | T | H | T | G | E | K | P | Y | V | C | R | E | C | G | - | G | F | I | Q | R | S | N | L | S | I | H | Q | R | T |
| JF750657.1\|Lycaon_pictus | T | H | T | G | E | N | P | Y | V | C | R | E | C | G | R | G | F | T | H | R | T | N | L | I | I | H | Q | R | T | H | T | G | E | K | P | Y | V | C | R | E | C | G | - | G | F | I | Q | R | S | N | L | S | I | H | Q | R | T |
| JF750658.1\|Canis_mesomelas | T | H | T | G | E | N | P | Y | V | C | R | E | C | G | R | D | F | T | H | R | T | N | L | I | I | H | Q | R | T | H | T | G | E | K | P | Y | V | C | R | E | C | G | - | G | F | I | Q | R | S | N | L | S | I | H | Q | R | T |
| JF750659.1\|Canis_aureus | T | H | T | G | E | N | P | Y | V | C | R | E | C | G | R | G | F | I | H | R | T | N | L | I | I | H | Q | R | T | H | T | G | E | K | P | Y | V | C | R | E | C | G | R | G | F | T | Q | R | S | T | L | N | E | H | Q | R | T |
| Island_Fox\|Urocyon littoralis | T | H | T | G | E | N | P | Y | V | C | R | E | C | G | R | G | F | T | H | R | T | N | L | I | I | H | Q | R | T | H | T | G | E | K | P | Y | V | C | R | E | C | R | * | G | F | T | - | - | - | - | - | - | - | - | - | - | - |
| Culpeo\|Lycalopex culpaeus | T | H | T | G | E | N | P | Y | V | C | R | E | C | G | R | G | F | - | H | R | T | N | L | I | I | H | Q | T | T | H | T | G | E | K | P | Y | V | C | R | E | C | G | R | G | F | T | Q | R | S | T | L | N | E | H | Q | R | T |

**Supplementary Table 6: Primers used for qPCR validation of ChIP**

| | |
|---|---|
| SYCE1.F | GTGTGTGGCATAAGAGTTTGTGTAT |
| SYCE1.R | AGATTAAAACAGGCAGGAGGAT |
| MLH1.F | TAGTGACCCAACTTAGTGTTTTCCT |
| MLH1.R | CTTTCTCTCCAGGTTCTTAACCTCT |
| Tex12.F | TATATGCACTCATGTCCCACTTAG |
| Tex12.R | AAACATCTACCCTTTCAGGATACAG |
| Tex26.F | TACTTGCTTGTCAGACTCTAGCA |
| Tex26.R | GGTATCTTTAGAGCCAATCTAGGAC |

All sequences are listed 5’ to 3’